%% file: JBH_Sharp_Edge_17.tex
\documentclass[iop]{emulateapj}
\usepackage{apjfonts}
\usepackage{graphicx}

\input{defs.tex}

\begin{document}

\title{In search of cool flow accretion onto galaxies - where does the disk gas end?}


\author{Joss Bland-Hawthorn}
\affil{Sydney Institute for Astronomy, School of Physics A28,
  University of Sydney, NSW 2006, Australia\\
  ARC Centre of Excellence for All Sky Astrophysics in 3 Dimensions (ASTRO-3D)} 
\author{Philip R. Maloney}
\affil{Centre for Astrophysics and Space Astronomy, University of Colorado, Boulder, CO 80309-0389, USA}
\author{Alex Stephens \& Anna Zovaro}
\affil{Sydney Institute for Astronomy, School of Physics A28,
  University of Sydney, NSW 2006, Australia}
\author{Attila Popping}
\affil{Centre of Excellence for All-sky Astrophysics, International Centre for Radio Astronomy Research, 7 Fairway, University of Western Australia, Crawley, Perth, WA 6009, Australia}
\begin{abstract}

The processes taking place in the outermost reaches of spiral disks (the
`proto-disk') are intimately connected to the build-up of mass and angular
momentum in galaxies. The thinness of spiral disks suggests that the activity is
mostly quiescent and presumably this region is fed by cool flows coming into the
halo from the intergalactic medium. While there is abundant evidence for the
presence of a circumgalactic medium (CGM) around disk galaxies as traced by
quasar absorption lines, it has been very difficult to connect this material to
the outer gas disk.  This has been a very difficult transition region to
explore because baryon tracers are hard to observe. In particular, \HI\ disks
have been argued to truncate at a critical column density (\NH $\gta 10^{19.5}$
cm$^{-2}$ at 30 kpc for an $L_\star$ galaxy) where the gas is vulnerable to the
external ionizing background. But new deep observations of nearby
$L_\star$ spirals (e.g. Milky Way, NGC 2997) suggest that \HI\ disks may
extend much further than recognised to date, up to 60 kpc at \NH\ $\approx$
10$^{18}$ cm$^{-2}$. Motivated by these observations, here we show that a clumpy
outer disk of dense clouds or cloudlets is potentially detectable to much larger radii and
lower \HI\ column densities than previously discussed.
This extended proto-disk component is likely to explain some of the
\MgII\ forest seen in quasar spectra as judged from absorption-line column
densities and kinematics. We fully anticipate that the armada of new radio
facilities and planned \HI\ surveys coming online will detect this extreme outer
disk (scree) material. We also propose a variant on the successful `Dragonfly'
technique to go after the very weak \Ha\ signals expected in the proto-disk
region.
\end{abstract}

\keywords{galaxies: individual (Milky Way, NGC 253, NGC 891, NGC 2997, NGC 3198, NGC 6946, UGC 7321), galaxies: ISM, evolution, intergalactic medium, quasars: absorption lines, radiative transfer}
     
\section{Introduction}
The details of how gas gets into galaxies remain something of a mystery. Does
it come in hot or cold, with or without dark matter, in a mildly coherent or a
highly turbulent process, at a constant rate or in a stochastic fashion, and so
on. The answers are intimately connected to how galaxies form and evolve over
cosmic time, and how baryons work with dark matter to build up angular momentum
over billions of years. The slow progress in understanding accretion is due to
the difficulty of observing gas in the outer reaches of galaxies. The hot, warm
and cold phases are all hard to observe directly in most galaxies beyond the
extent of the visible disk or halo.

Over the past decade, a theoretical picture has emerged where gas accreting into
galaxies is thermally and dynamically cold below a critical halo mass threshold
($\sim 10^{12}$\Msun) while more massive systems are fed by relatively hot gas
(e.g. Birnboim \& Dekel 2003; Kere\v{s} et al 2005, 2009). Most of this activity
occurred at $z=1-5$ although accretion is ongoing at a lower level for perhaps
half of all galaxies to the present day, including our own Milky Way. Building
on these ideas, the {\sl Horizon-AGN} simulations present a picture where mass
and angular momentum build-up in galaxies is a highly coherent process
(e.g. Dubois et al 2012; Codis et al 2015). The flow of gas down filaments with
velocity $\vec{v}$ leads to a net vorticity $\vec{\omega}$ ($=\nabla \times
\vec{v}$) that assists the build-up of angular momentum during collapse. Depending on the impact parameter of the
infalling gas, the vorticity arises partly
from oblique shocks as the gas
gravitates towards and along the dark-matter dominated filaments. The derived vorticity
today\footnote{The size of the vorticity is likely to scale inversely as $(1+z)$ and was therefore weaker in the past.} ($\omega \approx 100$ km s$^{-1}$ Mpc$^{-1}$)
is considerably larger than what is generated in a turbulent accretion
process (e.g. Cornuault et al 2016). 

\smallskip
\noindent{\sl The extent of the outer gas disk.}
Observational evidence for the cool flow scenario is suggestive but indirectly inferred for the most
part, e.g. from statistical
studies of QSO absorption-lines. Cool gas
traced by Ly$\alpha$ absorption has been detected ubiquitously around galaxies indicating the 
presence of a circumgalactic medium out to 150 
kpc in passive and star forming galaxies
(e.g. Tumlinson et al 2013). The mass of the CGM 
can be comparable to or exceed the stellar mass
(Faerman, Sternberg \& McKee 2017), and the strength of the Ly$\alpha$ absorption increases
towards the centre of the galaxy (q.v. Borthakur et al 2015). But are we seeing halo gas, a 
disk-halo transition region, or cooling material
settling to an outer `proto-disk'? The answer
is unclear although statistical evidence is 
emerging of a rotating torus-like region
in the outermost parts of disk galaxies (q.v. Ho et al 2017).


The most direct probe of the atomic hydrogen that typically dominates the outer
disks of $z=0$ spiral galaxies is the 21-cm \HI\ emission line.  Until recently,
the deepest \HI\ observations have rarely dipped below \NHI\ $\approx$
$10^{19.3}$ cm$^{-2}$ at $5\sigma$ in a narrow velocity channel (e.g. Westmeier,
Koribalski \& Braun 2013). In an important unpublished experiment (see van
Gorkom 1991), van Albada et al observed the isolated spiral galaxy NGC 3198 for
100 hours with the VLA in order to reach a column density sensitivity of $\NHI
\approx 4\times 10^{18}$ cm$^{-2}$, with the expectation of probing the
\HI\ emission to beyond $R\sim 60$ kpc. To their surprise\footnote{They would
  have been less surprised if the prescient paper of Bochkarev \& Sunyaev (1977)
  had not been unjustly overlooked.}, the $5\times 10^{19}$ cm$^{-2}$ and
$4\times 10^{18}$ cm$^{-2}$ contours were coincident within
their $1'$ beam (2.7 kpc for an assumed distance of 9.4 Mpc, but see below).

\smallskip
\noindent{\sl The impact of a cosmic ionizing background.} Our interpretation of ``cool'' gas
in the  extreme outer regions of galaxies is complicated
by the highly uncertain (and controversial)
cosmic ionizing background (CIB).
Following earlier discussions of the influence of 
the CIB on neutral hydrogen in galaxy disks (Sunyaev 1969; Felten \& Bergeron 1969; Bochkarev \& Sunyaev 1977), Maloney (1993) showed
that there should be a critical column density $N_c$ in
\HI\ disks at which the gas would go from largely neutral
to largely ionized, that $N_c$ would not depend strongly on either the galaxy
parameters or the intensity of the ionizing background, and that this transition
could plausibly explain the \HI\ truncation observed in NGC 3198, provided the CIB is of order $\phi_i\sim 10^4$ phot cm$^{-2}$ s$^{-1}$.

This idea was subsequently discussed by many authors with somewhat
inconclusive results (Corbelli \& Salpeter 1993; Bland-Hawthorn et al 1997;
Walsh et al 1997; Madsen et al 2001; Christlein et al 2010; 
Adams et al 2011;
Abramova 2012), and the observational situation in particular has 
not advanced significantly, largely because of the difficulty of achieving such
sensitivity and the requisite dynamic range. But \HI\ has now been observed down
to \NHI\ $\approx$ 10$^{18}$ cm$^{-2}$ revealing partial bridges, discrete
clouds, and extended gas disks around nearby disk galaxies in loose groups
(Braun \& Thilker 2004; Westmeier et al 2005; Pisano 2014; Wolfe et al 2013,
2016). Of particular note, Pisano (2014) used the Green Bank Telescope (GBT) to
reach down to \NH\ $\approx$ $1.2\times 10^{18}$ cm$^{-2}$ (5$\sigma$) in two
disk galaxies, NGC 6946 and NGC 2997. Although the single-dish GBT has a large
beam (a width of 16 and 32 kpc in NGC 6946 and NGC 2997, respectively), it does
not suffer from the spatial filtering of interferometer observations that make
them insensitive to emission on large angular scales. NGC 6946 was found to have
a previously undetected filament connecting it to its companion galaxies, while
NGC 2997 exhibited an extended \HI\ disk with an outer radius $R\approx 53$ kpc. By
analogy with the outer \HI\ disk simulations of Popping et al (2009), Pisano
suggested the compact filament could be part of a cool coherent flow (although a
tidal interaction origin is also plausible). These
papers have inspired our new calculations because the prospect of an extreme outer disk as a generic feature of most galaxies has major implications
for understanding how gas gets into disk galaxies.


\begin{figure*}[tbp]
\centering
\includegraphics[scale=0.45]{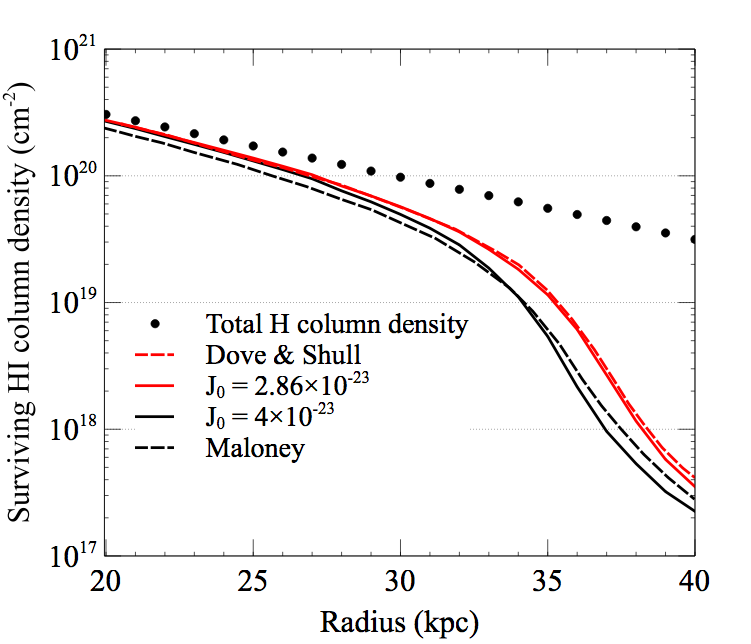}
\includegraphics[scale=0.45]{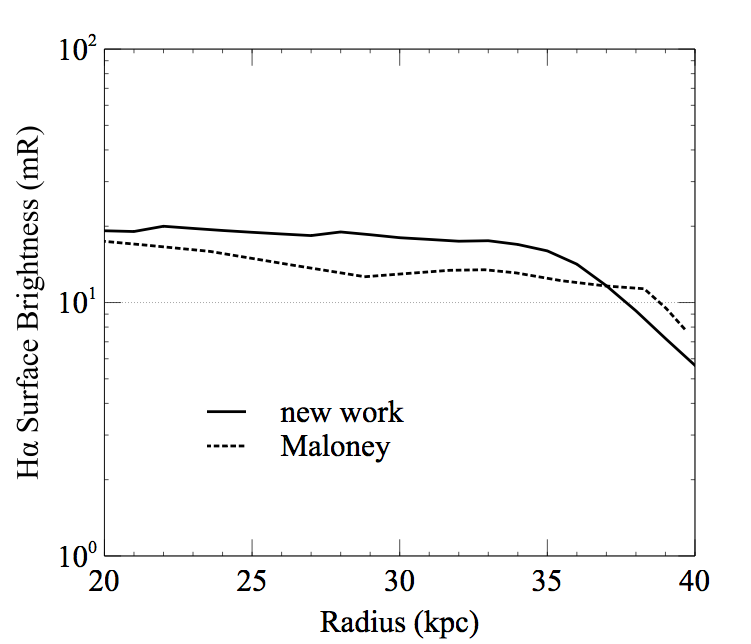}
\caption{(Left) Our computed curve of the dependence of column density \NH\ with
  galactic radius (solid red line) for comparison with Dove \& Shull (1994;
  dashed red line); we show the same comparison with Maloney (1993; dashed black
  line) and our model (solid black line). The adopted values of $J_o$ (in units
  of erg cm$^{-2}$ s$^{-1}$ Hz$^{-1}$ sr$^{-1}$) are chosen to match the
  respective values used by the above authors. The dots indicate an exponential
  profile for the neutral hydrogen emission in NGC 3198 prior to external
  photoionization.  The profile is shown over the limited radial range of 20 kpc
  to 40 kpc, as first presented by the above authors. The gas filling fraction
  and covering fraction are everywhere $f=1$, $c=1$ respectively.  (Right) Our
  computed \Ha\ surface brightness \Em\ as a function of galactic radius from a
  face-on disk (solid line) for comparison with Maloney (1993; dotted); this was
  not computed by Dove \& Shull (1994).  }
\label{f:compare}
\end{figure*}

\begin{figure*}[tbp]
\centering
\includegraphics[scale=0.45]{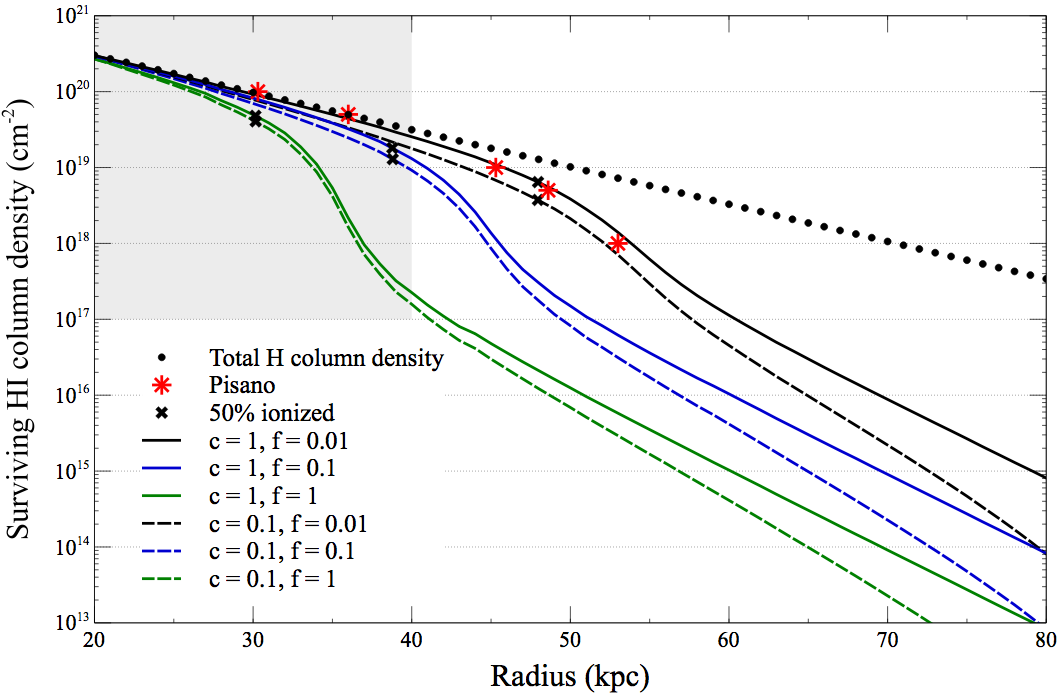}
\caption{The impact of different filling factors ($f$) and covering fractions ($c$) on the survival of neutral gas in the presence of external photoionization. The external ionizing field is the same as in Fig.~\ref{f:compare} with $J_o$ taken from 
Maloney (1993). The red asterisks are from Pisano (2014); the black
crosses indicate where the total \HI\ column is 50 percent ionized.
The solid and dashed curves (black, blue, green) correspond to
$f=0.01, 0.1, 1$ respectively as indicated. The dashed lines have matching fill factors but show the trend in decreasing covering fraction, from $c=1$ at 20 kpc to $c=0.1$ at 80 kpc. Higher values of $f$ increase the local gas density and make it more robust against being fully ionized. The black dots and green solid curve are the same as in Fig.~\ref{f:compare}, as shown by the shaded region; note that the radial scale is increased {\it threefold}. For guidance, \HI\ column densities down to $\sim 10^{16}$ cm$^{-2}$ can be detected in emission using radio receivers; QSO absorption lines reach down to much lower columns. The grey box shows the domain of the original study by Maloney (1993) in Fig. 1.}
\label{f:strong}
\end{figure*}


\begin{figure*}[tbp]
\centering
\includegraphics[scale=0.45]{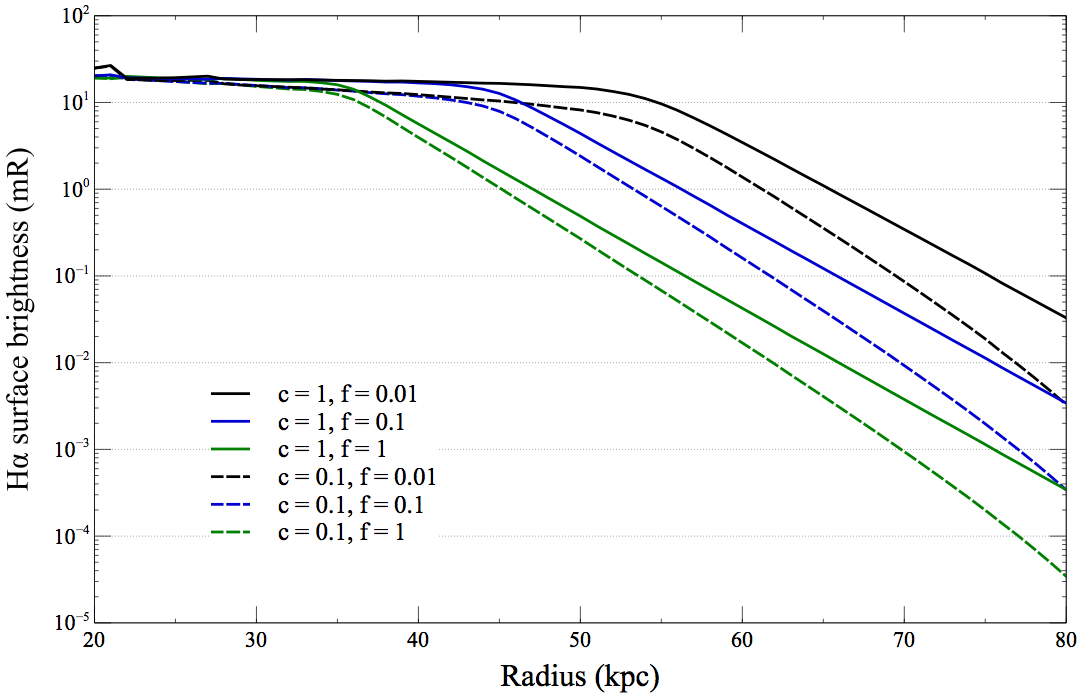}
\caption{The predicted \Ha\ surface brightness for the models presented in
  Fig.~\ref{f:strong} where the disk is seen face-on. The calculation is for
  two-sided ionization given that the cosmic ionizing intensity illuminates both
  sides of the \HI\ disk; the internal dust extinction is assumed to be
  negligible as expected in the outer parts of galaxies. For a disk inclined
  with angle $i$ (where $i=0$ is face on), the predicted values should be
  increased by $\sec i$ for $i < 90\deg$; see text for a full discussion on this
  correction. For guidance, an \Ha\ surface brightness below 1 mR is currently
  undetectable.}
\label{f:em_strong}
\end{figure*}

\smallskip
\noindent{\sl Inferring the cosmic ionizing intensity.}
We briefly review
the broad range of estimates for the CIB intensity 
to emphasize that its value is highly uncertain.
The present-day background ionizing intensity depends on 
environment and cosmic evolution
(\'Cirkovi\'c et al 1999; Maloney \& Bland-Hawthorn 1999, 2001; Haardt \& Madau
2012). Both of these contributions are only crudely constrained at the present
time. Weymann et al (2001) set a $2\sigma$ upper limit of 8 mR\footnote{1 milliRayleigh (mR) is defined
  as $10^3/4\pi$ \Ha\ phot cm$^{-2}$ s$^{-1}$ sr$^{-1}$ equivalent to $5.7\times
  10^{-21}$ erg cm$^{-2}$ s$^{-1}$ arcsec$^{-2}$ for an unresolved emission
  line.} for the
\Ha\ intensity of an opaque cloud illuminated by the CIB, although the
uncertainties are dominated by the sky subtraction and the uncertain geometry of
the cloud. Compared to Maloney's (1993) value
$J_o=4\times 10^{-23}$ erg cm$^{-2}$ s$^{-1}$ Hz$^{-1}$ sr$^{-1}$, and the $J=0.7 J_o$ value adopted by 
Dove \& Shull (1994), 
this is equivalent to an intensity at the Lyman limit of $J=0.4 J_o$. The theoretical estimate of the
present-day background of Haardt \& Madau (2012) is about a factor of two below
this upper limit ($J=0.2 J_o$).

In terms of the H ionization rate, the CIB intensity favored
by Maloney (1993) corresponds to $\Gamma_{\rm H}\approx 10^{-13}$ s$^{-1}$, while the
$z=0$ prediction of Madau \& Haardt (2015) is $\Gamma_{\rm H} = 2.3\times
10^{-14}$ s$^{-1}$
(cf. Fig. 11 of Fumagalli et al 2017). However, a number of papers on the low-redshift Ly$\alpha$
forest argued that their number was too small by a factor of a few (Kollmeier
et al 2014; Shull et al 2015; Khaire \& Srianand 2015; Viel et al 2017), and
the revised model of Madau \& Haardt (2015), updated with new quasar
emissivities, predicts a $z=0$ ionization rate of $\Gamma_{\rm H}\simeq 6\times
10^{-14}$ s$^{-1}$.

Very recently, Fumagalli et al (2017) have claimed detection of CIB-induced
\Ha\ fluorescence from the edge of the HI disk of the edge-on spiral NGC 7321.
Performing photoionization modeling of the disk\footnote{As in Maloney (1993),
  they assume a smooth distribution for the gas, and although they solve for the
  thermal as well as ionization structure of the gas, the gas scale height $z_g$
  is independent of the temperature, although they vary $z_g$ from an initial
  guess to match the \HI\ and \Ha\ observations.} they arrive at an H ionization
rate $\Gamma_{\rm H} \sim (6-8) \times 10^{-14}$ s$^{-1}$, which is in good
agreement with the theoretical and low-$z$ Ly$\alpha$ forest-derived numbers
discussed above. It is, however, in considerable disagreement with the upper
limit (and subsequent claimed $7\sigma$ detection\footnote{see
  http://iactalks.iac.es/talks/view/393}) of Adams et al (2011), which they
report as $\Gamma_{\rm H} = 2.0\times 10^{-14}$ s$^{-1}$ for the same galaxy,
even though the Adams et al (2011) upper limit to the \Ha\ surface brightness is
fully consistent with the detection of Fumagalli et al (2017). This should be
taken as an indication of the importance of careful photoionization modeling in
extracting $\phi_i$ from the \Ha\ observations, and that caution should be
exercised in accepting any particular value at present; as with Weymann et al
(2001), the uncertainty in the Fumagalli detection is largely systematic.

\smallskip
\noindent{\sl The main goals of our study.}
So how are we to understand the well
established existence of gas clouds $-$ even down to
\NH\ $\sim$ 10$^{17}$ cm$^{-2}$ (Wolfe et al 2016) $-$ far beyond the confines
of the opaque disk? Can we detect these clouds 
directly in future deep \HI\ and \Ha\ studies?
Do \HI\ truncations even exist at all?
{\it These are the main 
questions addressed by this study.}

If an extended (clumpy) component can be revealed as a common feature of disk
galaxies, we may be seeing the domain of the proto-disk fed by
cooling flows seen in some simulations (e.g. Stewart et al 2011).
We are led to revisit the key work of Maloney (1993) and the
corroborative study of Dove \& Shull (1994) which we extend to the
case  
of a clumpy medium. Given the uncertainty, we consider a broad (factor of ten) range in the value of $\phi_i$ to
cover the highest and lowest estimates of the present-day intensity of the
CIB and to facilitate comparisons with earlier work. Note, however, that the
critical column $N_c$ only depends on the square root of the extragalactic
ionizing flux (see below). As we show, present-day gas disks can in principle extend a factor of
two beyond the radial limit at \NH\ $\approx$ $10^{19.3}$ cm$^{-2}$ of earlier work, and may be detectable 
to these outer limits with new facilities under
development.


The structure of the paper is as follows. In \S 2, we briefly recapitulate the
major assumptions and results of Maloney (1993), and discuss what this model
does and does not imply for \HI\ distributions in disk galaxies, as there has
been some confusion in the literature on this point. In \S 3, we discuss our
galaxy model in detail before describing the computational method in \S 4. The
results of the calculations are discussed in \S 5. In \S 6, we look forward to
what will be possible in the next few years. 
At many scalelengths beyond the optical edge, disk stars
are rarely observed and this may conceivably be the domain of the coherent cold
flow. But to map these structures over large areas will require a different
observational approach and large amounts of dedicated telescope time, as we
discuss. In \S 7, we examine the broader implications of our work.



\section{The critical column density and disk \HI\ edges}

There are only limited theoretical studies that directly or indirectly 
infer the maximum radial extent of \HI\ in disk galaxies from 
first principles, i.e. in the absence of ionizing radiation.
Numerical simulations of $L_\star$ galaxies make a case for 
\HI\ settling to a disk out to at least 100 kpc in radius (Kere\v{s} \& Hernquist 2009; Stewart et al 2011; Nuza et al 2014), but these are often complex filamentary
structures.

Simple arguments can be made that an exponential \HI\ disk is a 
consequence of cooling in a spinning hot halo (Fall \& Efstathiou
1980; Mo, Mao \& White 2010). In support of this 
picture, the hot coronae ($\sim 10^6$ K) of several nearby high-mass disk galaxies have
been detected through either stacking or radial binning of a direct x-ray image
(Anderson \& Bregman 2011; Dai et al 2012; Miller \& Bregman 2013).  A disk condensing out of the hot
halo can extend to a
significant fraction of a virial radius if the hot halo has sufficient angular momentum. 
In these models, the start time for accretion
moves outwards from the centre, and the accretion timescale
increases as a function of the galactic radius.
Using this approach, Tepper-Garcia \& Bland-Hawthorn (2017) produce a dynamically stable model of the Galaxy
in line with the observed halo emissivity constraints. The
atomic hydrogen disk has a scalelength of $r_d \approx 7$ kpc and \NHI\ $\approx$ $3\times 10^{17}$ cm$^{-2}$ at a radius of $r = 80$ kpc.
This is the non-ionized hydrogen profile we 
adopt in this work which is broadly consistent with the
observed \HI\ profile in the Galaxy (see below;
Kalberla \& Dedes 2008).





Independent of any detailed modeling, Maloney (1993) showed that
the truncation radius at which \HI\ is observed
can be much less than the outer boundary of the gas disk. He presented a simple argument
for the critical atomic hydrogen column density in a galactic disk, 
to which the
gas goes from largely neutral to largely ionized, by equating the incident
ionizing photon flux to the total column recombination rate within the vertical
gas layer. This results in the expression
\bee 
N_c = 2\pi^{1/4}\left({\phi_i\; z_{g}\over \alpha_{\rm rec}}\right)^{1/2}
\eee
where $z_g$ is the scale height of the gas and $\alpha_{\rm rec}$ is the
recombination coefficient. This can be put into a more useful form by relating
$z_g$ to the galactic mass parameters, with the result (see Maloney 1993 for details)
\bee 
N_c \approx 2.6\times10^{19} \left({\phi_{i,4} \sigma_{g,10} V_{150}\over
  \Sigma_{h,100}}\right)^{1/2} {\rm cm^{-2}}
\eee
where $\phi_i=10^4 \phi_{i,4}$ phot cm$^{-2}$ s$^{-1}$, the gas vertical
velocity dispersion is $\sigma_g = 10 \sigma_{g,10}$ km s$^{-1}$, the asymptotic
value of the disk rotation velocity $V=150\; V_{150}$ km s$^{-1}$ and the halo
mass surface density (which is a function of radius) $\Sigma_h = 100
\Sigma_{h,100}$ \Msun\ pc$^{-2}$. 
This analytic estimate was shown to agree very
well with the detailed photoionization model results.\footnote{For the numerical
  results, the critical column was defined as the column where the neutral
  fraction dropped to 10\%. However, since the neutral fraction varies very
  rapidly around $N_c$, the precise definition of the neutral to total ratio for
  the numerical $N_c$ is not crucial.}

The most important assumptions about the
atomic gas that go into this calculation are that it is warm and smoothly
distributed in the galactic potential; $\alpha_B$ evaluated at $T=10^4$ K has
been used for the recombination coefficient. Because $V$ and the magnitude of
$\Sigma_h$ are tightly related, the ratio of these quantities does not vary
strongly from galaxy to galaxy or within galaxies over the radii of
interest. The vertical velocity dispersion is typically $\sigma_g = 6-10$ km
s$^{-1}$ for the bulk of the \HI\ observed in face-on disk galaxies and
independent of radius in the outer disks; this is also comparable to the
expected magnitude of $\sigma_g$ for gas in the warm neutral phase (but see the
discussion below). 

What does the existence of such a critical column density imply for galactic
\HI\ disks? Notice that this is predicted to be a generic feature of \HI\ disks,
provided the assumptions about the gas distribution hold, and that $N_c$ is not
strongly dependent on either $\phi_i$ or the galaxy parameters. Note also that even
if the (lower) Haardt \& Madau (2012) CIB intensity is correct, $N_c$ will drop by only
about a factor of two. For NGC 3198, Maloney (1993) argued that the observed
truncation could be explained by photoionization, provided that the total
hydrogen column continued to fall as extrapolated from the inner part of the
disk (where the total and neutral hydrogen columns are nearly identical).
Although reportedly the truncation in NGC 3198 is independent of azimuth (ruling
out a warp, for example, as an explanation), only the major axis cut was
available for analysis. There are noticeable differences in \HI\ extent between
the NE and SW sides of the disk, with the SW being more extended. However, such
asymmetries are not really surprising due to the long dynamical timescales at
these radii: the orbital period for gas at 30 kpc in NGC 3198 is more than
$10^9$ years. The Milky Way's \HI\ distribution, for example, is noticeably
asymmetric, with more gas in the south than in the north (Kalberla \& Dedes
2008). The observed neutral hydrogen distribution along the two sides of the
major axis was reproduced with a model in which the deviations on the two sides
from the same exponential fall-off in total H were less than 50\%.

Hence the moderate asymmetry between the NE and SW major axis cuts is not really
an issue for the photoionization model. However, it does illustrate an important
point: even if all of the model assumptions are satisfied, whether
photoionization by the CIB results in a sharp truncation of the disk depends on
the shape of the underlying total hydrogen distribution. If this exhibits
asymmetries or a change in behavior with increasing radius, this will be
reflected in the observed \HI\ distribution. A gas disk with a shallow radial
profile can exhibit an extended \HI\ disk even if the column density has dropped
below the critical column.

If the assumptions about the nature of the gas distribution are violated,
however, then the results can be very different. For example, in a deep
\HI\ observation of the edge-on spiral NGC 891 (which has a noticeably lopsided
\HI\ distribution between the two sides), Oosterloo et al (2007) found
that the radial extent of the gas increased very little from earlier studies,
but discovered a low column density component that extended to large distances
above the plane; they argued that this was evidence against the
CIB-photoionization model for the radial edges of the \HI\ disk, which were
likely to be physical edges. We note simply that (a) this is very plausible, but
(b) the presence of gas at such large heights above the plane indicates physical
conditions very different from those required for the models of Maloney (1993) to
apply, e.g., the presence of a galactic fountain, in which case the
high-altitude atomic hydrogen is almost certainly cooling and condensing and
unevenly distributed in a very clumpy medium.
In addition, the \HI\ disk of NGC 891 is much less
extended compared to the optical disk than that of NGC 3198: on the northern
side of the disk it only extends about 4.5 optical scalelengths, while for the
southern side it is about 6. Hence it is more likely that the
\HI\ distribution in NGC 891 has been disturbed by stellar processes; the very
extended \HI\ disk of NGC 3198 (more than 11 disk scalelengths) is precisely why it was chosen by van Albada et
al for their deep imaging experiment.

Similarly, Kalberla \& Dedes (2008) in their study of the Milky Way's
\HI\ emission found that gas extended to $R\sim 60$ kpc at low column densities,
and argued that this was inconsistent with the results of Maloney
(1993). However, the fit to their \HI\ data shows an abrupt shallowing of the
radial distribution beyond 35 kpc and the vertical velocity dispersion required
to explain the vertical extent is 74 km s$^{-1}$. Whatever the physical
mechanism responsible for this distribution, it is unavoidable that this gas is
highly clumped, as the velocity dispersion cannot possibly represent thermal
motion.

The distribution and structure of the gas in the outer disks of galaxies is a
complicated subject in which many uncertainties remain. The 21-cm
\HI\ emission-absorption study of the outer Milky Way --- the galaxy for which
we have the best data --- of Dickey et al (2009) showed that the mean spin
temperature of the atomic hydrogen is nearly constant out to 25 kpc in radius,
implying that the mix of cool and warm gas is similarly independent of both
radius and height above the disk.\footnote{A similar conclusion was reached by
  Curran et al (2016) for their sample of galaxies, which showed no evidence for
  radial variations in spin temperature over the inner $R\sim 30$ kpc of the
  galactic disks.} Roughly 15-20\% of the gas is in the cool phase, while the
rest is warm. This implies that changes in the interstellar environment (in
particular, the drop in average ISM pressure) with R and $z$-height do not
result in the disappearance of the cool phase. Both the cool and warm gas have
scale heights that increase with radius. Strasser et al (2007) showed that the
distribution of the cool gas in the outer Galaxy is not random, but is
preferentially found in large, coherent structures that may be connected to the
Galaxy's spiral arms, suggesting that it may be gravitational perturbations that
trigger the cooling of the cool atomic gas out of the warm neutral medium.

To study the detailed structure of the Milky Way's \HI, Kalberla et al (2016)
combined data from the Galactic Effelsberg-Bonn HI survey with the third release
of the Parkes Galactic All Sky Survey and, using an unsharp masking technique,
found an abundance of small-scale cold filaments that they argue show the CNM is
largely organized in sheets which are themselves embedded within a WNM with
approximately ten times the column density of the CNM sheets ($\NHI\sim 10^{19}$
cm$^{-2}$); sheets, rather than clumps, are what is expected when pressure
effects dominate over self-gravity. To what extent the outer \HI\ disk of the
Milky Way resembles the outer disk of a galaxy such as NGC 3198 is an open
question.

Even in an otherwise smooth distribution, the introduction of clumpiness into
the gas distribution can have dramatic impact on the resulting radial
\HI\ distribution in the CIB photoionization model. In what follows we include a
very simple prescription for clumping of the gas into the photoionization model
and show how this can lead to extended disks at much lower column densities than
in the unclumped models.



\section{Disk ionization model} 

\subsection{Introduction}

We summarize the framework of Maloney (1993) and Dove \& Shull (1994) before
introducing some modifications necessary for this work.  The work concentrates
on the $L_\star$ galaxies NGC 3198 and NGC 2997 as they share key properties
with the Milky Way (Maloney 1993; Pisano 2014).
They also share essentially identical distance moduli
placing them at about the same distance (Tully et al 2016).  The Cosmicflows-3
catalogue gives a distance modulus for NGC 3198 of 30.61$\pm$0.17 using both
Cepheid variables and the Tully-Fisher relation; for NGC 2997, the Tully-Fisher
distance modulus is for group member ESO 434-34 of 30.74$\pm$0.40 in close
proximity. From their respective rotation curves, NGC 3198 is less massive than
NGC 2997 by a small factor although their different inclinations (edge on
vs. face on) makes the factor uncertain. We ignore this difference below as the
corrected scaleheight will involve an even smaller factor. For continuity with
Maloney (1993) and Dove \& Shull (1994), we assume both galaxies have the same
radial scaling at a distance of 9.4 Mpc. {\it This scaling is 40\% smaller than
  implied by the new distance modulus from Cosmicflows-3. Thus there is a case
  for increasing all quoted radii throughout the paper by the same amount.} But
we refrain from such a large correction to maintain continuity with the earlier
work.

Both galaxies show gas disks that extend far beyond the optical emission. In NGC
3198, as discussed above, based on VLA observations, after a fairly smooth
decline the disk appears to truncate (somewhat unevenly between the two sides of
the major axis) at radii between 33 and 40 kpc, where the projected column
density $\NHI\sim 5\times 10^{19}$ cm$^{-2}$; note that corrected for the
galaxy's substantial inclination ($i\approx 72^\circ$), this is a face-on column
density $\NHI\sim 2\times 10^{19}$ cm$^{-2}$. In NGC 2997, which is at low
inclination and hence geometric corrections are small, the disk extends much
further out, to approximately 53 kpc at the $\NHI\sim 1.2\times 10^{18}$
cm$^{-2}$ level, and any steepening in the slope of \NHI\ versus radius occurs at or beyond the 
limits of the data.

Our model is based on the original observations of NGC 3198 (Maloney 1993)
although we extend the work to a discussion of both NGC 2997 and the Galaxy.
The galactic disk is modelled as having a spherical dark matter halo and
exponential stellar $+$ gas disk. Azimuthal symmetry is assumed, thus
restricting the problem to two dimensions in $(R,Z)$ where $R$ denotes
galactocentric radius and $Z$ vertical height from the midplane. Due to symmetry
about the midplane, the problem can be further restricted to $Z \geq 0$.  {\it
  We scale up the projected measures (e.g. \Ha\ surface brightness) for
  two-sided photoionization.} For the gas column
  densities relevant to the outermost \HI\ disk,
  the disk opacity is negligible for the expected
  metallicities.

\subsection{Galactic model} 

In our ionization model, the disk is divided into a series of radial (annular)
bins 1 kpc apart, going outwards from the galactic centre. If these radial bins
are insulated against radiation emitted by neighbouring bins, then the problem
is reduced to a one-dimensional problem in $Z$ at any radius $R$.  The mass
column density of stellar matter at a given radius is given by 
\bee 
\Sigma_\star(R)=\Sigma_\star(0)\exp\left(-\frac{R}{R_\star}\right)\text{ g
  cm\textsuperscript{-2} } 
\label{e:stellarCD}
\eee
where $R_\star$ is the disk scalelength, which we take to be 2.7 kpc in line
with the earlier work.  The column density of gas at a radius $R$ is given by
\bee
\Sigma_{gas}=\Sigma_0(1+4y)\exp\left(-\frac{R}{R_0}\right)\text{ g
  cm\textsuperscript{-2}} 
\label{e:gas+DMCD}
\eee
where $y$ is the abundance of helium relative to hydrogen and the gas scale
length is $R_0 = 8.8$ kpc (Maloney 1993). Note that $R_\star$ and $R_0$ are
broadly consistent with the Galaxy (see above).  The dark matter halo has a
density given by
\bee
\rho_{dm}(Z)=\frac{\rho_{dm,0}}{1+(R^2+Z^2)/a^2}
\eee
where $a$ is the core radius ($=$7.92 kpc) and $\rho_{dm,0}$ ($=6.7\times
10^{-3}$\Msun\ pc$^{-3}$) is the density at the galaxy's centre (Dove \& Shull
1994).

The galactic model is also explored for a disk with different filling factors
$f$ to account for a non-uniform density within each column; this increases the
local density and makes the gas more difficult to ionise. We concentrate on
$f=(0.01, 0.1, 1)$ which covers the range of reasonable atomic hydrogen
densities\footnote{The use of $\langle\rangle$ denotes volume averaged densities which are less than local densities if $f<1$; all other density variables refer to local densities.} ($\nH=0.1-10$ cm$^{-3}$). Because the possible ways of implementing
such a volume filling factor are practically unconstrained, and because our main
focus is on illustrating the effects of such clumpiness, we have chosen an
extremely simple form for $f$.

We also explore the impact of a non-uniform
covering fraction of gas $c$ such that $c<1$ indicates there are holes in the
gas disk when viewed face-on. Clumpiness is seen in simulations
of the outermost reaches of \HI\ disks (e.g. Stewart et al 2011).
The covering fraction is a 2D (projected) quantity;
the factors $f$ and $c$ can both be
operating simultaneously and are entirely independent in our treatment. Thus,
for $c<1$, we do not attempt to conserve the gas mass at each radius. The
necessary modifications are given in the next section.


\begin{figure*}[tbp]
\centering
\includegraphics[scale=0.45]{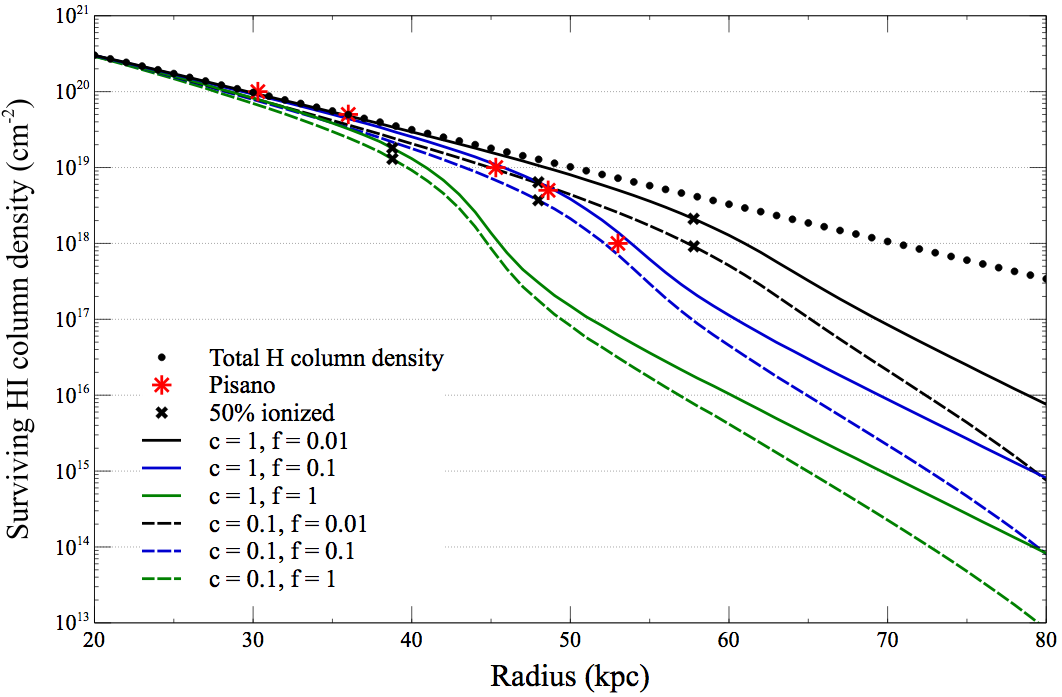}
\caption{The impact of different filling factors ($f$) and covering fractions
  ($c$) on the survival of neutral gas in the presence of external
  photoionization. The external ionizing field is now $J=0.1 J_o$, i.e. 10\% of
  the assumed cosmic intensity in Fig.~\ref{f:strong}. The solid curves correspond to $f=0.01, 0.1, 1$ as indicated. The dashed
  lines have matching fill factors but show the trend in decreasing covering
  fraction, from $c=1$ at 20 kpc to $c=0.1$ at 80 kpc. Higher values of $f$
  increase the local gas density and make it more robust against being fully
  ionized.}
\label{f:weak}
\end{figure*}


\begin{figure*}[tbp]
\centering
\includegraphics[scale=0.45]{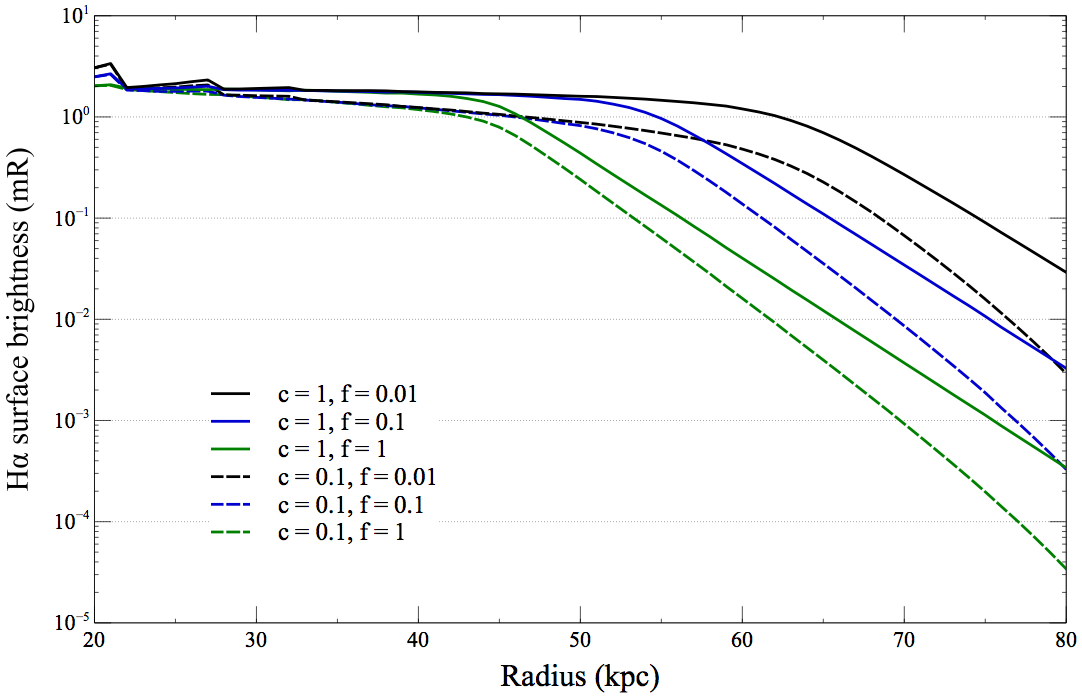}
\caption{The predicted \Ha\ surface brightness for the models presented in
  Fig.~\ref{f:weak} where the disk is seen face-on. The calculation is for
  two-sided ionization given that the cosmic ionizing intensity illuminates both
  sides of the \HI\ disk; the internal dust extinction is assumed to be
  negligible as expected in the outer parts of galaxies. For a disk inclined
  with angle $i$ (where $i=0$ is face on), the predicted values should be
  increased by $\sec i$ for $i\lesssim 90\deg$. For guidance, an \Ha\ surface
  brightness below a few mR is currently undetectable.}
\label{f:em_weak}
\end{figure*}

\subsection{Vertical gas distribution}

The column density of \HI\ gas at a radius $R$ is given by eq.~\ref{e:gas+DMCD}. The
equation is highly idealized and may not describe the typical galaxy in detail. For
example, in the Galaxy, a double exponential is needed to trace the gas
(e.g. Kalberla \& Dedes 2008). But our model is simple because it is grossly
modified by the external radiation field.

In our calculation, we wish to find the mass density of gas $\rho$ at a given
height $Z$. Since the local disk potential in the outer disk is dominated by the dark matter,
this is given by the equation of
hydrostatic equilibrium
\bee
\sigma_{zz}^2\frac{\partial \rho(Z)}{\partial Z}=-\rho(Z)K_z
\label{drhodz}
\eee
where $\sigma_{zz}$ refers to the vertical velocity dispersion ($\lesssim 10$ km s$^{-1}$) and $K_z$ the
acceleration due to gravity, coupled with Poisson's equation 
\bee
\frac{dK_z}{dZ}=-4\pi G \rho^{\prime}(Z)
\label{dkdz}
\eee
assuming isothermality of the disk at T = $10^4$ K.
Again, this
is only a working approximation because other poorly constrained
sources (e.g. magnetic fields, cosmic rays) may contribute (Cox 2005).
The density
$\rho^{\prime}$ now represents the local mass density in toto (stars, gas, dark matter). For our 1D calculation, we include the filling factor by decreasing the calculated gas scale height by $f$ ($f \leq 1$). This compresses the gas distribution, raising the midplane density by a factor of $1/f$. This simplified treatment is intended only to illustrate how increasing the local gas densities from the smooth case will affect the truncation and extent of the disk.

A more sophisticated analysis would be to treat radiation transfer in 3D
through a porous (fractal) medium (e.g. Bland-Hawthorn et al 2015). Such a framework would be much more realistic, but also introduces a level of poorly-constrained complexity that is not justified for the presently available data. We return to this point in \S 6.


In our adopted simple
form, the change in the critical column density in the presence of a volume
filling factor $f$ can be accounted for by inserting $f$ in the numerator of
equations (1) and (2), so that $N_c$ scales like $f^{1/2}$. Two models were used: a constant fill fraction $f$ for all $R$, and a linear
model in which $c=1$ below a given cutoff radius, and then decreases linearly
to its minimum value at $R_{\rm max}$. The idea
of a clumpy medium may appear to be inconsistent
with equations~\ref{drhodz} and \ref{dkdz}. 
But here we assume that the clumpy
medium is confined by a hot diffuse
phase transparent to radiation with negligible mass.

\subsection{Radiation field}

The specific intensity of the isotropic extragalactic radiation field is assumed
to obey a simple power law: 
\bee
J_\nu = J_o \left({{\nu}\over{\nu_o}}\right)^{-\alpha}
\label{e:cosmic}
\eee
where $\nu_o$ is the frequency of the Lyman limit, $J_o$ is the specific
intensity of radiation (in $ \text{ergs cm\textsuperscript{-2}
  s\textsuperscript{-1} Hz\textsuperscript{-1} sr\textsuperscript{-1}}$) at the
Lyman limit and $\alpha\approx 1$ for a spectrum dominated by AGN.

\section{Computational method}
\paragraph{Gas layers:}
Our method closely follows the approach of Dove \& Shull (1994).  For a given
$R$, the vertical gas density profile is computed as follows: first, an initial
estimate for the midplane density $\rho_0$ is assumed.
This is used in a fourth-order Runge-Kutta routine to find $\rho(Z)$ and $K_z$
using equations \ref{drhodz} and \ref{dkdz}; $\rho(Z)$ is then integrated over
$Z$ to find the column density. This procedure is iterated until the column
density converges to that found using equation~\ref{e:gas+DMCD}.  This method is
used to divide the disk into a number of layers in $Z$, where each layer has a
constant gas density.
We decided on $m=60$ layers as this was found to yield sufficient accuracy; fewer
layers led to stochastic variations at high \NH.

\subsection{Photoionisation and Recombination Equilibrium}

Photoionisation and recombination equilibrium is satisfied when the number of
ionisations is equal to the number of recombinations per unit volume per unit
time. For an optically thick hydrogen nebula, this is given by (Osterbrock \&
Ferland 2006) 
\bee
n_{\HI}\int_{\nu_0}^{\infty}\frac{4\pi J_{\nu}}{h\nu}\sigma_{\nu}d\nu = 
n_e n_p\alpha_{\HI}^{B}(T)
\label{nebular}
\eee
where $J_{\nu}$ is the angle-averaged specific intensity of incident radiation,
$\sigma_{\nu}$ is the frequency-dependent photoionisation cross-sectional area
of hydrogen and $\alpha^{B}_{\HI}$ is the recombination coefficient of hydrogen
in an optically thick medium. This equilibrium is more generally expressed by
the following equality between photoionisation and recombination rates:
\bee
X_{i,j}\Gamma_{ij}=n_eX_{i,j+1}\alpha_{ij}(T)
\eee
where $X_{i,j}$ is the fraction of species $i$ in ionisation state $j$,
$\alpha_{ij}$ is the Case A recombination rate coefficient for
electron captures into the $j$th ionization state,
and $\Gamma_{ij}$ denotes the total ionisation rate for species $i$
into state $j+1$, given by
\bee
\Gamma_{ij}=\int_{\nu_{ij,0}}^{\infty}\frac{4\pi
  J_{\nu}}{h\nu}\sigma_{\nu,ij}d\nu + \Gamma^{sec}_i 
\eee
where, for species $i$ in ionisation state $j$, $\sigma_{\nu,ij}$ is the
photoionisation cross-sectional area and $\nu_{ij,0}$ is the minimum frequency
of a photon which can ionise the species. $\Gamma^{sec}_i$ refers to the
ionisation rate of secondary ionisations, which are those resulting from
high-energy electrons ionising neutral hydrogen. However, this component is
negligible and can be ignored for the given electron temperature of $T_e=10^4$ K
(Dove \& Shull 1994). 

There are two sources of ionisation in the galactic disk --- a {\it direct} and a
{\it diffuse} component. The direct component refers to that from the isotropic
cosmic background, and the diffuse component refers to the photons emitted by
recombinations, namely those with sufficient energy to ionise hydrogen. 

\paragraph{\textit{Direct component.}}
The direct component resulting from the isotropic cosmic background radiation,
$J^{\rm dir}_{\nu}$, at a layer in the disk, is given by 
\bee
J^{\rm dir}_{\nu}=\frac{1}{2}\int_{-1}^{0}I_{\nu}e^{\tau_{\nu}/\mu}d\mu
\label{Jdir}
\eee
where $\tau_v$ is the optical depth between the layer and the point at which the
radiation enters the disk, and $\mu$ is the cosine of the angle between the
normal to the layer and the line of propagation into the layer (Maloney 1993). We have assumed that there are no additional sources of radiation in the extreme outer disk.

\paragraph{\textit{Diffuse component.}}
After Maloney (1993), the diffuse mean specific intensity incident on one layer
due to another layer, at height Z, an optical depth $\tau_{\nu}$ away, $J^{\rm
  diff}_{\nu}$, is given by
\bee
J^{\rm diff}_{\nu} = \frac{1}{2}S_{\nu}(Z) \int_{0}^{1}
[e^{-\tau_{\nu}/\mu}-e^{-(d\tau_{\nu}+\tau_{\nu})/\mu}] d\mu
\label{Jdiff}
\eee
where $\tau_{\nu}$ is the optical depth between the layers and $d\tau_{\nu}$ is
the optical thickness of the source layer at $Z$. $S_{\nu}(Z)$ is the source
function at height $Z$, simply given by the recombination rate 
\bee
S_{\nu}(Z)=n_en_{i,j}\alpha_{ij}(T) \text { phot cm\textsuperscript{-2}
  s\textsuperscript{-1}} 
\label{sourcefunc}
\eee

The total rate of photoionisation is influenced by both Eq.~\ref{Jdir}
and Eq.~\ref{Jdiff}. However, as they are interdependent --- linked by electron
density --- an iterative scheme is required to numerically compute the final
ionisation fractions in each layer. Taking into account the different species
present in the galactic model, the total electron density is  
\bee
n_e = n_{\text{H II}}+n_{\text{He II}}+2n_{\text{He III}}
\eee
Therefore the resulting fraction of an ionised species is given by 
\bee
\frac{X_{i,j+1}}{X_{i,j}} =
(n_e\alpha_{ij}(T_e))^{-1}
\int_{\nu_{ij}}^{\infty}
\frac{4\pi J_{\nu}}{h\nu}\sigma_{\nu,ij}\;d\nu
\label{ionfrac}
\eee
where $J_{\nu} = J^{\rm diff}_{\nu} + J^{\rm dir}_{\nu}$.

The computational method comprises two phases: the initial calculation of the
ionisation state from the direct radiation component only, using the
`on-the-spot' approximation to treat the effect of recombinations (e.g.,
Osterbrock \& Ferland 2006), followed by a more accurate iterative equilibrium
computation.

After Maloney (1993), there are several sources contributing to the diffuse
intensity $J_{\nu}^{\rm diff}$. These include
\begin{itemize}
\item Ground-state recombinations of \HI, \HeII\ and \HeIII\
\item The $2^3S\rightarrow 1^1S$ (19.8 eV) and $2^1P\rightarrow 1^1S$ (21.3 eV)
  resonance-line transitions of helium 
\item The $2^1S \rightarrow 1^1S$ (two photons, combined energy 20.6 eV)
  transition of \HeI\ 
\item The $2^2P\rightarrow 1^2S$ (40.7 eV) Lyman-$\alpha$ \HeII\ transition
\item \HeII\ Balmer-continuum emissions ($E\geq$13.6 eV)
\item The $2^2S \rightarrow 1^2S$ (two photons, combined energy 40.7 eV)
  transition of \HeII\ 
\end{itemize}
Case A recombination coefficients 
are used in this step, as it is no longer assumed that all recombinations
produce photons which are absorbed locally, and so ground-state recombinations
must be considered in calculations (Osterbrock \& Ferland 2006). 

An initial guess is taken for the hydrogen column density at a given radius $R$,
and used to compute a new value until the two converge, giving the final value. 
For each of the diffuse-radiation producing processes, the following routine is
carried out, beginning with initial guesses for species densities:  
the photoionisation cross-sectional area is calculated.
Optical depths for the relevant frequency are calculated using
$d\tau_{\nu}=\sigma_{ij}n_{i,j}\Delta l$, $l$ denoting layer height, which are
then used to find the recombination rate in each layer using
eqn. \ref{sourcefunc}, using the previously calculated species densities. Then
the diffuse radiation resulting from this process incident on each layer,
including contributions from all other layers in the disk (including those on
the other side of the midplane), is calculated using equation \ref{Jdiff}. 

After this has been carried out for each process, the resulting diffuse
radiation intensity values are used in eqn. \ref{ionfrac} to compute the
ionisation fraction of the species in each layer. \HII, \HeII\, and
\HeIII\ number and column densities can then be computed. The whole process is repeated using these as new starting
values, until the values converge to no more than a $10^{-3}$ discrepancy. The
final electron density is used to calculate the emission measure $f n_e^2\Delta
l$ where $n_e$ is the electron density in each of $m$ layers and $\Delta l$ is
layer height, and where $f$ is the filling factor. This process is repeated for
each radial block.

\subsection{Emission measure}
\label{ss:emissionMeasure}
An emission measure is a useful physical quantity for relating the ionising
photon flux to the induced H$\alpha$ surface brightness.  It has a long history
in atmospheric physics and low surface brightness astronomical observations; for
an extended discussion, see Tepper-Garcia et al (2015). 
This quantity is
calculated at each radial value using the simple formula for an ionized nebula taken from Spitzer (1978)
\bee
\Em(\Ha)= \int_{0}^{L}n_e n_p dl \;\;\;\; {\rm cm^{-6}\; pc}
\eee
effectively giving the sum of the total H recombinations along the line of sight
$L$ where $n_p$ is the proton number density. For a plasma at temperature $10^4$ K, an emission measure of 1 cm$^{-6}$ pc
is equivalent to an \Ha\ surface brightness $\mu(\Ha) = 330$ mR.

\subsection{Comparison with earlier work}
In Fig.~\ref{f:compare}, we show our computed radial variation in \NH\ and
$\mu$(\Ha) with earlier work.  Our computational method is closer to Dove \& Shull
(1994) and thus the agreement for \NH\ is excellent when using the same cosmic
ionizing intensity.  These authors do not compute the trend in \Em\ or equivalently $\mu$(\Ha).  When we
rescale the CIB upwards to match Maloney (1993), the
agreement is not as good for the run of \NH\ but is sufficient for our purposes;
the same holds for the trend in $\mu$(\Ha).

\bigskip
\input{NH-cutoff-table.tex}


\begin{table*}
\begin{center}
\caption{Fields of view (FoV), angular and spatial resolutions, $1\sigma$ \HI\ column densities (\NH), \HI\ mass sensitivities and survey speeds (SS) for different telescopes based on a total integration time of 100 hours. For all telescopes, this is the time to map an area 
the size of the SKA-MID FoV (0.43 deg$^2$), or larger if indicated. All physical quantities are scaled to a distance of 10 Mpc. Footnotes: {\it a}: based on Popping et al (2015), {\it b}: Nan et al (2011), {\it c}: extrapolated from Wolfe et al (2016), {\it d}: Heald et al. 2009, {\it e}: http://imagine-survey.org, {\it f}: http://mhongoose.astron.nl}
\begin{tabular}{lccccccc}
\hline
{\bf Telescope} & FoV$_\eta$ & Res. & Res. & \NH\ & Mass & SS & SS$_{\NH}$ \\
                  & deg$^2$ & [arcsec] & [$(D/10$ Mpc) kpc] &  [cm$^{-2}$  20 km/s] & $[(D/10$ Mpc$)^2$\Msun] & & \\
\hline
SKA-MID$^a$ & 0.43 & 29 & 1.4 & 1.5e17 &  5.0e3 & 1.0e-4 & 0.085 \\
MeerKAT$^a$ & 0.66 & 30 & 1.4 & 4.1e17 & 1.5e4  &  1.7e-5 & 0.015 \\
ASKAP$^a$ & 24 & 21 & 1.0 & 8.8e18 & 1.6e5 & 5.4e-5 &  0.0024 \\
JVLA-C$^a$ & 0.16 & 25 & 1.2 & 2.7e18 & 6.5e4 & 4.8e-6 & 0.003 \\
JVLA-D$^a$ & 0.16 & 83 & 4.0 & 2.5e17 & 6.5e4 & 4.8e-6 & 0.033 \\
FAST$^b$ & 0.021 & 180 & 8.5 & 2.1e16 & 1.4e4 & 5.3e-6 & 0.17 \\
GBT$^c$ & 0.01 & 550 & 27 &  2e16 & 1.3e5 & 4.7e-8 & 0.014 \\
\hline
{\bf Survey} & & & & & \\
\hline
HALOGAS$^d$ (WSRT) & 0.16 & 40 & 2.0 & 2.5e18 & 8.5e4 & &\\
IMAGINE$^e$  (ATCA) & 0.20 & 140 & 7.0 & 2.4e17 & 1.0e5 & &\\
MHONGOOSE$^f$  (MeerKAT )& 0.66 & 30 & 1.4 & 1.2e18 & 4.8e4  & & \\
MHONGOOSE$^f$  (MeerKAT) & 0.66 & 90 & 3.5 & 1.3e17 & 4.8e4   & &\\
\end{tabular}
\label{t:HI}
\end{center}
\end{table*}

\section{Results}

In relatively isolated or loose-group spiral galaxies, the \HI\ disk typically
extends $1.5-2$ times further than the optical radius ($\mu_V=25$ mag
arcsec$^{-2}$) at a limiting column density of \NH\ $\sim$ 10$^{20}$ cm$^{-2}$
$-$ for a recent review, see Bosma (2016). These outer regions can be complicated
by warps and flares, or even asymmetries induced by the galaxy's motion with
respect to the local medium (e.g. Heald et al 2016). We have ignored these
details and concentrated on the survival of cool gas in the presence of an
external radiation field.

Our interest has been kindled by recent discoveries of low column \HI\ gas at
very large radius from the parent galaxy (Pisano 2013; Wang et al 2016). One of
the more spectacular examples is seen in the outskirts of M31 which exhibits a collection
of \HI\ clouds at a radius of $\sim$90 kpc in the direction of M33. The clouds
have sizes of order 1 kpc and masses of order 10$^5$\Msun\ observed at
\HI\ column densities below 10$^{18}$ cm$^{-2}$. For the Galaxy, Kalberla \&
Dedes (2008) fit the \HI\ column density profile with a double exponential
function which extends to at least 60 kpc down to \NH\ $\approx$ $10^{18}$
cm$^{-2}$. In comparison, other large galaxies like M83 have huge \HI\ disks
that extend beyond 80 kpc in radius at a $10\times$ higher column density limit
of 10$^{19}$ cm$^{-2}$ (Koribalski 2017). At present, one can only wonder just how far this gas
extends at lower column density. Our new work is intended to motivate the
community to explore this tantalizing question.

We add a further note of caution: our predicted radial 
profiles for the \HI\ column density (\NH) and
the expected \Ha\ surface brightness ($\mu$(\Ha))
assume a smooth, face-on spiral disk.
Observed profiles are normally `boosted' by the 
effects of disk inclination. {\it For smooth,
diffuse emission, the observed profiles must
be deprojected before a direct comparison can be
made with our models (see below).} If the emission
is patchy, the situation is more complex and
will require a more careful treatment.

\medskip
\noindent{\sl Derived column densities.}  In Figs.~\ref{f:strong} and
\ref{f:weak}, we revisit the calculations of Maloney (1993) and Dove \& Shull
(1994) over a much larger radial range. We also consider the geometric filling
and covering factors $f$ and $c$ which allow us to examine the impact of clumpy
gas in a consistent and controlled manner. As the local density rises through
decreasing $f$, the gas is more able to survive photoionization at a fixed $J_o$
through faster recombination. The smallest fill factor we consider is $f=0.01$
as this produces clouds with densities at the upper range expected in outer
disks ($\sim$10 H atoms cm$^{-3}$).

It is important to note that there is {\it still} a critical column density
present in all of the models: it has simply moved to lower column densities (and
hence to larger radii) as a consequence of the filling factor; there is also a
secondary effect of the transition getting shallower with decreasing $f$.


As discussed in Maloney (1993), the gas in NGC 3198 truncates at a radius of
about 33 kpc to the NE. This behaviour follows the $c=1 $, $f=1$ (solid green)
track closely in Fig.~\ref{f:strong} ($J=J_o$). But such a cut-off is much less
evident on the opposite side of the galaxy, where the gas extends
(non-monotonically) to 40 kpc. As discussed above, this may simply be a
consequence of modest asymmetries in the total hydrogen distribution within the
disk, which are common in spiral disks and are not terribly surprising at
large radii. Another possibility is that the asymmetry is the result of
differences not just in total column density but in the degree of clumpiness of
the atomic gas between the two extremes of the major axis.

Alternatively, it is possible the \HI\ disk is shaped at least partly by some
other process, such as ram pressure (although NGC 3198 is notably isolated and
the asymmetries of the disk are quite modest). The NE side may be confined by an
external medium; the SW side may represent the outer gas becoming clumpy. 
This type of strong to mild asymmetric ram pressure confinement of the outer
\HI\ disk is seen in dozens of nearby disks, including NGC 7421 (Ryder et al
1997), NGC 2276 (Gruendl et al 1993), the Large Magellanic Cloud (Kim et al 2003; Salem et al 2015) and in the highly-inclined Sculptor galaxy NGC 253 (Koribalski et al 1995).

Pisano (2014) does not provide direct information (e.g., major axis cuts) of the
radial \HI\ distribution in NGC 2997 because of
the low resolution of the GBT observations. However, his Table 1 provides the extent
(area) of the \HI\ disk as a function of column density. We have used these data
to derive the red points plotted in Fig.~\ref{f:strong}; in effect, this is the
symmetrized radial \HI\ distribution for the modestly asymmetric disk.  The
plotted data match very well with the $c=1$, $f=0.01$ model. We note that at
some level this agreement is fortuitous: the mass model used has not been
matched to the rotation curve of NGC 2997 (although the effects, assuming a flat
rotation curve in this region, will be modest), and we have not attempted to
match the \HI\ distribution at smaller radii --- we note also that the large beam of
the GBT means that the inner disk is not well resolved. However, it does
demonstrate that such a model can easily explain the extent of the disk and the
slow roll-off of the neutral hydrogen.  Now consider Fig.~\ref{f:weak} where we
present the same model parameters for a weaker cosmic ionizing intensity ($J=0.1
J_o$) at the low end of the likely CIB intensity. Again fortuitously, the red
points fall along the $c=1$, $f=0.1$ (solid blue) track. This demonstrates the
strong degeneracy between $J$ and $f$. As the models show, there is a degeneracy
between $J$, $f$ and $c$.

In our simple model for NGC 3198 (Fig.~\ref{f:strong}), the \HI\ column
densities are (10$^{20}$, 10$^{19}$, 10$^{18}$, 10$^{17}$, 10$^{16}$) cm$^{-2}$ at (30, 50,
70, 90, 110) kpc before photoionization by the CIB. After imposing a high value of $J_o$, the
crossing points plummet to (27, 34, 37, 42, 51) kpc
(see Table~\ref{t:NH}). But for a gas-rich system with a covering fraction of unity, the filling factor has
a major impact on the extent of the detectable \HI\ gas. Even for our strongest external radiation field, the crossing points move out to
(29, 46, 54, 61, 70) kpc for the case $f=0.01$. The 50\% ionization radius now occurs at
$r_{0.5} = 48$ kpc, i.e. the neutral extent of the
disk has more than doubled in area. The crossing points for all models
are presented in Table~\ref{t:NH}.

In Fig.~\ref{f:weak}, we consider a weak external 
field, $J=0.1 J_o$.  
For $f=1$, the crossing points are now approximately (29, 41, 46, 51, 60) kpc as a function of the column density. For our lowest $f$ value, these move out to
(30, 49, 61, 69, 79) kpc.
The 50\% ionization radius now occurs at
$r_{0.5} = 58$ kpc such that we may
expect to find detectable \HI\ gas {\it twice} as far out as indicated by the early
NGC 3198 data, at least in principle.

For all models, the columnar ionization fraction is increasing with radius. In
the last row of Table~\ref{t:NH} we list the radius at which the total column of
hydrogen becomes 50\% ionized. As indicated, the 50\% ionization radius for
Maloney (1993) is $r_{0.5} \approx 30$ kpc at a column density of \HI\ above
10$^{19.5}$ cm$^{-2}$. For all other models, $r_{0.5}$ occurs at lower \NH\ and
is up to a factor of two further out in the most extreme cases.

A broad distribution in $f$ is a natural consequence of a turbulent medium
(Lazarian \& Pogosyan 2000). A spread in $c$ can arise from increased clumping of the
outer disk gas as the cold-flow accretion channel becomes more dispersed, as
observed in numerical simulations (Nuza et al 2014; Scannapieco et al 2015;
Crain et al 2017). Thus we consider the effect of slowly declining
covering fraction as shown by the dashed lines in Figs.~\ref{f:strong} and
\ref{f:weak}.  As expected, a correction for a decreasing covering fraction
shortens the crossing point distances.

Patchy gas makes detection more difficult generally.  This can lead to both a
lower covering fraction within the telescope beam, and a complex line profile in
velocity space. Simply stated, a covering fraction of $c$ requires a telescope
with an aperture that is $1/c$ times larger to observe the same signal under the
same observing conditions. The deepest HI observations require the largest
apertures if the gas fills or partially fills the beam. But if the gas is in the
form of dense `nuggets,' higher sensitivity can be achieved by resolving out the
clumps, e.g. spectroscopy with a very long baseline interferometer (VLBI).

\smallskip
\noindent{\sl Derived emission measures.} In Figs.~\ref{f:em_strong} and
\ref{f:em_weak}, we also present the predicted \Ha\ surface brightnesses for the
strong and weak external field, respectively. The boosted signal from clumping, if relevant, is included in predictions. For the maximum
cosmic intensity, we expect 20 mR at best, declining to a few mR at the limit of
the data; for the weak external field, the expected value is an order of
magnitude less.  As already mentioned, an \Ha\ surface brightness of 1 mR is
exceedingly faint, corresponding to $5.7\times 10^{-21}$ erg cm$^{-2}$ s$^{-1}$
arcsec$^{-2}$ in more conventional units (Bland-Hawthorn et al 1995).  This
level of sensitivity is an order of magnitude fainter than what is possible in
conventional spectroscopy and needs some explanation. There are two approaches,
the first which average over most or all of the detector to achieve the
detection limit, and the second from stacking many independent spectra.

Bland-Hawthorn et al (1997) use Fabry-Perot `staring' where the detector is
azimuthally binned to produce a narrow spectroscopic band ($\sim
10$\AA). Apertures are typically in the range 5 arcminutes to 1 degree
(e.g. Madsen et al 2001). The smaller apertures are well matched to the
resolution of the biggest single-dish radio telescopes. Zhang et al (2016) use a
very different method. They identify 0.5 million galaxies with 7 million
sight-line spectra which are co-added once shifted to a laboratory rest frame.
They detect the presence of emission line gas out to 100 kpc and identify this
as being primarily from the hot halo. In reality, a substantial contribution may
come from an extended thick disk.

In principle, it is possible to boost the projected signal from the cosmic
ionizing intensity. The one-sided ionizing photon rate (phot cm$^{-2}$
s$^{-1}$) is given by
\bee
\phi_i = \pi \int_{\nu_o}^{\infty} {{J_\nu}\over{h\nu}}\; d\nu
\eee
where the cosmic ionizing spectrum $J_\nu$ is defined in eq.~\ref{e:cosmic}.
This reduces to $J_o=\alpha h \phi_i/\pi$ which establishes a simple
relationship between $\phi_i$ and $J_o$, both of which can be related easily
to the \Ha\ surface brightness $\mu$, i.e.
\bee
\mu(\Ha) = 4.5 {\cal G} \left({{\phi_i}\over{10^4\; {\rm phot\; cm^{-2}\;
      s^{-1}}}}\right)\;\;\;\; {\rm mR} 
\label{e:em}
\eee
assuming the \HI\ slab of gas is optically thick to UV radiation (Bland-Hawthorn
\& Maloney 1999).  In fact, we can do better this as indicated by the geometric
factor ${\cal G}$.  Weymann et al (2001) note that the high inclination $i$ of a
flat structure of gas (\HI\ 1225+01) can be exploited because of the increased
pathlength through its ionized outer surface. There is an additional factor of
two from the gas sheet being ionized on both sides. Thus, under certain
circumstances, we can boost the observed signal in eq.\ref{e:em} by ${\cal
  G}=2/\cos i$. This factor cannot increase indefinitely and is likely to break
down for angles steeper than about $i=70\deg-75\deg$.  The ${\cal G}$ values for
NGC 3198 ($i=72\deg$) and NGC 2997 ($i=30\deg$) are $\lta$5.9 and 2.3
respectively.  This method allowed Weymann to set the deepest outer disk upper
limit to date, i.e. 8 mR ($2\sigma$). A higher
de-projected upper limit of about 20 mR was
established from 
null detections in the extreme outer disk of M31 (Madsen et al 2001; cf. Adams et al 2011).

For individual galaxies, the deepest outer disk detection is from
Bland-Hawthorn et al (1997).
These signals at 50$-$100 mR are suprisingly strong, even after
considering geometric corrections, and reflect either the presence of star
formation activity at exceedingly low levels, or the warp of the outer disk
causing the gas to be exposed to the inner radiation field, i.e. self-ionization
(Bland-Hawthorn et al 1998).  Interestingly, there is new evidence of very weak
star formation in the outer warped disk of one disk galaxy which makes it possible to
age-date the onset of the warp (Radburn-Smith et al 2016). Such activity may be affecting outer disk 
measurements in other galaxies (Christlein et al 2010; Fumagalli et al 2017).

The recent claim of a roughly 20 mR indirect detection of the CIB in the outer disk of UGC 3721 is not obviously 
consistent with the non-detections mentioned above,
especially the Weymann et al (2001) result,
even with due consideration for patchy emission
or de-projection (Fumagalli et al 2017).
The deepest non-detection experiments were undertaken
with Fabry-Perots in `staring' mode 
(Bland-Hawthorn et al 1994, 1995).
In principle, the much larger ${\cal R}\Omega$ product of a Fabry-Perot interferometer,
where ${\cal R}$ is the spectral resolving power
and $\Omega$ is the instrument solid angle, 
makes it far more sensitive
to diffuse light than conventional spectrographs.

However, the Fumagalli result is consistent with estimates for the CIB intensity based on the low-redshift Ly alpha forest and the most recent theoretical estimates of the background, as discussed in the Introduction. Given these disparate results, caution in interpreting current measurements is mandatory, but as we discuss in the next section, there is some prospect of a more definitive detection.



\section{In search of the outer disk material}

\noindent{\sl Faint} \Ha\ {\sl emission.}
So what are the prospects for
detecting this outer proto-disk material in spatially resolved imaging or
spectroscopy? We stress that detailed models will
be needed to associate the beam-smoothed \HI\ and
\Ha\ 
measurements discussed here with the intrinsically
much higher spatial resolution observations from
QSO absorption line measurements.

The predicted emission measures for the warm ionized gas are
likely to be too low for most galaxy environs, as we have shown. Even in the
most advantageous situations, we need stacking over many galaxies or the
Fabry-Perot `staring' technique which washes out the structure. 
Encouragingly, in the past few years, the first detections of emission from warm gas in the circumgalactic environment ($R\sim 100$ kpc) have been made.
In all cases, the surface brightness is exceedingly faint, requiring long
integration times, data stacking, or areal binning in observations with our most
advanced telescopes. 
Zhang et al (2016) detect \Ha$+$[NII] emission at
exceedingly faint levels ($\sim 3$ mR) in the $ 50-100$ kpc radius bin in 7 million stacked spectra from the
SDSS survey. This is an order of magnitude fainter than the levels observed by
Bland-Hawthorn et al (1997) and Christlein et al (2010) in their survey of
\Ha\ emission at or beyond the \HI\ edge of nearby galaxies, although the
detection limit of the former method is of order 1 mR at 1$\sigma$
(Bland-Hawthorn et al 1994; Madsen et al 2001).

The Dragonfly
technique (van Dokkum, Abraham \& Merritt 2014) of using many Canon cameras with
their near-perfect anti-reflection coatings is worth pursuing, but this approach
is hampered by the need for a narrowband ($\lta 10$\AA) filter. These cannot be
installed in the fast camera beam (Bland-Hawthorn et al 2001) and instead must
be inserted at the entrance aperture of the telescope.\footnote{This method was
  used successfully by Parker \& Bland-Hawthorn (1995) for the UK Schmidt
  \Ha\ survey with an entrance \Ha\ filter (10\AA\ FWHM) 14 inches in
  diameter. The crude optics and detector system meant that the detection limit
  was $\gta$300 mR.}  
  
In order to exclude faint starlight, the resultant images
must be compared with broadband images using a red filter presumably from a
companion Dragonfly system. The combined signal could reasonably get down to a few mR smoothed over a
10\arcsec\ patch of sky for one or more fields in several dark nights. This is the
sensitivity needed if we are to detect the ionized layer in the proto-disk region.
Our estimate comes from the 3$\sigma$ broadband
($B$, $I$)
limit (30 mag arcsec$^{-2}$) of the Dragonfly system.
Here we assume this performance holds for the $R$
band. For background limited observations (sky$+$galaxy
continuum), an emission-line region with a surface brightness of 500 mR detected within the band ($\Delta\lambda_R$ $\approx$ 140 nm) is detected with the 
same sensitivity.  If we replace the 
$R$ filter with a 100$\times$ narrower bandpass filter, as in the
UKST experiment, our detection limit falls to 50 mR.
We get another factor of 10 gain (5 mR at 3$\sigma$) by smoothing the diffuse data over a 10\arcsec\ patch.
Our wild extrapolations assume the instrument response (e.g. flatfielding) can be removed to better than 0.1\% or so (Bland-Hawthorn et al 1994, 1995). 
 


\smallskip
\noindent{\sl Faint \HI\ emission.}  If the goal
is to `image' the proto-disk region, we argue here that
the best prospect is to go after the clumpy cool
gas emission. Support for this idea comes from the
simulations of Nuza et al (2014) where clumpy \HI\
gas is "observed" to occupy a complex, roughly planar network of filaments - we refer to this scree of material as the
{\sl proto-disk region}. Their
simulations are broadly tuned to the characteristics of the Local Group. The cool proto-disks for M31 and the
Galaxy are seen to overlap and to extend out to at least 100 kpc from both galaxies.\footnote{In a comment on this work, K.C. Freeman posed an interesting question as to whether the filamentary
\HI\ networks in groups like Leo and M81, long assumed to be tidal debris from outer disk
interactions, are more likely to arise from interlocking proto-disk regions. The distinction
here is that the outer disks are only now forming
for the first time.}

Interestingly, Wolfe et
al. (2013, 2016) imaged an area of 12 square degrees mapping the region
between M31 and M33 with a $5\sigma$ detection level of \NH$=3.9\times10^{17}$
cm$^{-2}$ over 30 km s$^{-1}$ and found discrete clouds that have a typical size
of a few kpc and an \HI\ mass of $10^5$\Msun.  Upcoming observations of nearby
galaxies will achieve this sensitivity or better (Table~\ref{t:HI}). Popping et
al (2009) gave detection limits for a number of telescopes including ASKAP and
the SKA using the best sensitivity estimates at the time. Here we expand this
table with better sensitivity predictions and include contemporary surveys as a
point of reference.

Popping et al (2015) have made detailed performance simulations of the SKA and
SKA precursors based on the actual baseline configuration and telescope
design. We use their methods to derive new sensitivity estimates based on the
latest specifications as currently presented by the SKA office. In
Table~\ref{t:HI}, we present $1\sigma$ column densities and \HI\ mass
sensitivities for a number of different facilities. The mass estimates come from
assuming the limiting column density is uniform across the beam.  In order to
compare the performance of different radio telescopes, both interferometers and
single dish telescopes, we make certain assumptions. In our analysis, we assume
that the area imaged corresponds to a field of view of 0.43 deg$^2$ if the quoted field is smaller
than this solid angle. This
is chosen as it is equivalent to the field of view of SKA-MID which sets the benchmark for future studies.
Single dish telescopes can only
observe a single pointing equivalent to their angular resolution and therefore
we take into account the number of pointings that are required to map the same
solid angle.

As our aim is to search for low column density \HI\ gas around galaxies, we
assume an observation time of 100 hours per object. This integration time is
typical for these kind of studies, e.g. HALOGAS on WSRT (Heald et al 2010),
IMAGINE on ATCA (http://imagine-survey.org) and MHONGOOSE on MeerKAT
(http://mhongoose.astron.nl). The angular resolution of interferometers is
dependent on the baseline distribution and moderated by the weighting and
tapering of the visibility data. In order to make a fair comparison, we use the
natural resolution of each interferometer $-$ without using any weighting $-$ as
this gives the highest flux sensitivity and represents the resolution that is
best matched to the baseline distribution.  In Table~\ref{t:HI}, the columns and
cloud masses are ideal for targets like NGC 3198 and NGC 2997. The tabulated
numbers are scaled to a distance of 10 Mpc for consistency with both galaxies.

To understand which radio telescope is best for mapping an extended area that is larger than a single pointing, we calculate the survey speed defined as SS $=$ FoV$_\eta / $rms$^2/t_{\rm int}$ where FoV$_\eta$ is the effective field of view and $t_{\rm int}$ is the integration time. The effective field of view represents the noise equivalent area of the beam and is given by FoV$_{\eta} = (\pi / 8)(1.3\lambda / D)^2$. The survey speed is relevant when concentrating on the flux sensitivity, but becomes meaningless for the surface brightness which depends on the (synthesised) beam area. To encapsulate this difference, we define a surface brightness survey speed as SS$_{\NH} =  ({\rm FoV}_{\eta} \cdot b_{maj} \cdot b_{min}) / $rms$^2/t_{\rm int}$. For interferometers, we calculate the flux and brightness sensitivities without any tapering of the visibility data and using natural weighting (no down-weighting of data). This will give the best flux sensitivity at the natural resolution. The synthesised beam of an interferometer can be altered by weighting of the UV data, which can greatly enhance the surface brightness sensitivity. For ASKAP and FAST, the effective field of view is larger than a single beam due to the phased array feed on ASKAP and the 19-multibeam on FAST. The extraordinary potential of the recently commissioned FAST telescope is clear. Another powerful approach is \HI\ absorption-line spectroscopy which favours interferometers observing bright background sources. This experiment will be addressed in future work.

\smallskip
\noindent{\sl Combined \HI\ and \Ha\ studies.}
As we have discussed, there is a degeneracy
between the intensity of the CIB and the total hydrogen distribution that
produces an observed neutral hydrogen distribution. This also depends on the
degree of clumping of the gas and on the gas covering factor. One way to break
this degeneracy is through simultaneous measurements of both the \HI\ and
\Ha\ emission. For a smooth gas distribution, this is discussed in detail
by Fumagalli et al (2017), who used the combination of the \HI\ profile with their
measured \Ha\ flux to derive their constraints on the CIB.

Detailed mapping of both the neutral and ionized emission potentially offers a
powerful probe of both the CIB intensity and the neutral gas column density
distribution and the degree of clumping; provided that the same gas dominates
both the \HI\ and \Ha\ emission, any patchiness of the disk emission will
drop out. We have not discussed such models for two main reasons:
(i) It would be very premature. At present the needed datasets simply do not
exist, and although we have suggested ways in which they might be obtained ---
and for \HI\ emission the needed facilities are either now or soon to be available
--- at the moment, increasing the complexity of the models
cannot be justified. Note that the detailed modelling carried out by Fumagalli to constrain the CIB assumed a smooth distribution of gas, and essentially
was based on just two observed quantities, the location of the ionization front
(what we refer to in this paper as the truncation radius) and the surface
brightness of the \Ha\ emission interior to the truncation radius. (ii) For a clumpy disk, it is not only necessary to adopt a 3D treatment of the radiative
transfer but assumptions also need to be made about the gas clumpiness as a
function of radius and height above the plane. Even in the Milky Way, these are
not well-determined quantities, and hence model parameters that are poorly
constrained would be introduced. What should drive this type of modelling would
be the discovery from high S/N data sets (at minimum, well-resolved radial
profiles of both the \HI\ and \Ha\ emission) that no reasonable smooth models
were capable of simultaneously fitting both the \HI\ and \Ha\ data.


\section{Discussion}

Outer disk regions remain mysterious. Most disk galaxies appear to be warped and often flared in the outermost parts,
at least in \HI\ (Briggs 1990). Another recent 
revelation is that the declining metallicity gradient
of both stars and gas appears to flatten out beyond several disk scalelengths (q.v. Vlajic 2010; Sanchez et al 2014). It is
interesting to consider whether some of the extended \HI\ envelopes identified
in nearby groups (e.g., Leo, Cen, M81, M83) normally attributed to tidal debris
from high column regions is in fact compressed material from proto-disk
regions. How can we distinguish between these cases? This is something that future simulations might address. 

But it remains unclear how many scalelengths in radius we need to explore to
establish clear evidence of cool flows from the intergalactic medium. Just how
the metal-poor gas arrives, and in what state, is a topic of debate. 
With reference to the QSO absorption line work,
are we seeing halo gas, a 
disk-halo transition region, or cooling material
settling to an outer `proto-disk' ?
Curran et al (2016)
combining data from a number of surveys and incorporating upper limits, 
found the first evidence for an
anti-correlation between \HI\ 21-cm absorption strength and impact parameter in
galaxies. This suggests that there may be a condensation process at
work in which the galactic ISM condenses from the CGM, which in turn is accreted
from the intergalactic medium. 

Borthakur et al (2015) studied the ISM-CGM
relation in a sample of \HI-detected galaxies from the GASS survey whose
circumgalactic media were probed via COS Ly $\alpha$ absorption in the spectra
of background quasars with impact parameters in
the range 60-230 kpc. Roughly 90\% of the
galaxies showed Ly$\alpha$ absorption, and there 
was a strong correlation
between the galaxy atomic gas mass fraction $(M_{\HI}/M_*)$ and the impact
parameter-corrected Ly$\alpha$ equivalent width, suggesting a physical
connection between the atomic ISM and the CGM, i.e., the \HI\ disks are being
fueled by accretion from the CGM. Studies of the kinematics of low-redshift Mg
II absorbers, from modeling of line profiles, suggest a combination of disk
rotation plus radial infall and/or radial inflow (Charlton \& Churchill 1998; Ho
et al 2016).

Arguably the most tantalizing evidence for some kind of accretion flow comes
from the outer reaches of redshifted galaxies probed by quasar sightlines
(e.g. Charlton \& Churchill 1998; for a recent review, see Ho et al 2017).  The
\MgII\ absorption doublet in the UV has long been known to trace the \HI\ gas
down to \NH\ $\sim$ $10^{17}$ cm$^{-2}$ in disk galaxies, i.e.,  gas that is
optically thick at the Lyman limit (Bergeron \& Stasinska 1986; Steidel et al
2002).

From an extensive literature, a `typical' redshifted disk galaxy produces the
following absorption signatures: damped \Lya\ ($R\lesssim 15$ kpc);
\MgII\ ($R\lesssim 100$ kpc); \OVI\ ($R\lesssim 200$ kpc). The damped
\Lya\ limit is from Charlton \& Churchill (1998); the \MgII\ limit is from Ho et
al (2017); the \OVI\ limit is from Kacprzak et al (2015). The radial limits
scale statistically with a number of other factors: (i) the galaxy's $K$-band
luminosity (Steidel et al 1995); (ii) the cosmic ionizing intensity: at lower
redshift, \MgII\ can be traced to larger radius because the ionizing intensity
is weaker (Bergeron et al 1994).

The trend in recent papers is to ascribe some of the action in QSO sight lines
to a spherical outflow (Nielsen et al 2017), particularly at high redshift. The
observed epoch is critical: what we infer during the golden age of star
formation and AGN activitity ($z=1-5$) does not necessarily follow for all
time. In contrast to the strong kinematically broadened structure at earlier
times, co-planar gas is clearly seen in the weak absorbers (Kacprzak \&
Churchill 2011; Kacprzak et al 2011; Matejek et al 2013; Kacprzak et al 2015; Ho et al 2017).

Tumlinson et al (2013) provide a census of the neutral hydrogen absorption in the halos of a sample of 44 low-redshift ($z \sim 0.2$) $L \sim L_{\star}$ galaxies with the Cosmic Origins Spectrograph on HST, in which the QSO sightline passes within 150 kpc of the galaxy (the COS-Halo sample). Of most relevance to this paper, they find (i) strong HI absorption in the circumgalactic media of their sample galaxies, (ii) that this absorption arises at velocities indicating that the absorbing material is bound to the galaxies, and (iii) this absorption arises in gas at temperatures much lower than the virial temperature, indicating that the gas has either cooled and condensed subsequent to shock-heating or was never heated to the virial temperature. Comparison with other samples of Ly$\alpha$ absorption indicates that there is an increase in the strength of Ly$\alpha$ absorption within an impact parameter of $\sim$200 kpc, consistent with the kinematic association of the absorption with the galaxies in the COS-Halo sample.

In principle, these data could be used to constrain the column density distributions we discuss here. However, in practice there are difficulties, some of which Tumlinson et al discuss in their own comparison with galactic disks (their \S 6.3.1). Foremost among these is the very large uncertainty in the COS-Halo \HI\ column densities over the range at which they would provide the most constraints. Due to saturation of the line profiles, all of the COS-Halo measurements within an impact parameter of 70 kpc are lower limits (aside from one damped Ly$\alpha$ system), with the caveat that the column densities are probably below 
\NH\ $\approx$ $10^{18-18.5}$ cm$^{-2}$, 
due to the absence of damping wings in the line profiles. Hence all of the COS-Halo \NHI\ columns have a one to three order of magnitude uncertainty over the precise range of impact parameter (20 to 70 kpc) in which measurements would be most useful.

What we can say is that the Tumlinson results are inconsistent with a model in which their QSO sightlines intercept disks such as we model here if the areal filling factor is close to unity at R $\sim$ 40 kpc and below. Given the very small number of galaxies with impact parameters this small in their sample, this is not so surprising (see also their comparison with the Kalberla \& Dedes Milky Way disk model). We note that areal filling factors of 0.3-0.5 quoted in their paper have modest impact on beam-averaged emission models compared to their effects on absorption statistics.

Binney (1992) concluded in a review on outer disks that `one's best guess must be that warps {\it will} in the end prove to be valuable probes of
cosmic infall and galaxy formation.' Ultimately, there {\it must} be a co-planar `proto-disk' region if we are to
account for the thinness of disks (Matthews, van Gallagher \& van Driel 1998), their
exponential profiles (Mestel 1963; Fall \& Efstathiou 1980) and their outer warps (Briggs
1990; Binney 1992). Some have suggested that this co-planar region is the
likely site of inflow accretion (e.g. Kacprzak et al 2015) but clear kinematic
evidence of this is hard to come by at present. The spatially resolved
structure and extent of this proto-disk region is unknown. The numerical
simulations (e.g. Popping et al 2009; Nuza et al 2014; Scannapieco et al 2015)
are fairly rudimentary at this stage and need to include more microphysics and
higher spatial resolution before convergence is achieved across different
hydrodynamical codes (Kere\v{s} et al 2012).

We have made a case that, under favourable conditions, the proto-disk region may
be directly detectable in emission. Powerful radio facilities
(Table~\ref{t:HI}) are coming on line which can directly tackle this question
and we urge telescope allocation committees to embrace the unknown. New
facilities should always be pushed to their limits; short time allocations
rarely break new ground. If successful, this would herald a new era in tracing
how disk galaxies are fed by the intergalactic medium.

\section{Acknowledgment}
JBH is supported by a Laureate Fellowship from the Australian Research Council. AS is supported under this grant. AZ was supported by JBH's earlier
Federation Fellowship from the Australian Research Council. JBH acknowledges
insightful conversations with Ken Freeman, B\"arbel Koribalski and Jacqueline van Gorkom over two decades on the nature of the outer disks of spirals. We are grateful to Yuval Birnboim, 
Naomi McClure-Griffiths, Jay Lockman, John Dickey 
and Ron Ekers
for continued inspiration in this wonderfully
rich and underappreciated field. We are indebted to an inspired
referee for their deep insight and assistance.


\end{document}

%% file: defs.tex
\def\apj {ApJ}
\def\apjl {ApJL}
\def\apjs {ApJS}
\def\aj {AJ}
\def\mnras {MNRAS}
\def\aap {A\&A}
\def\nat {Nat}
\def\araa {ARAA}

\def\bea{\begin{eqnarray}}
\def\eea{\end{eqnarray}}
\def\bee{\begin{equation}}
\def\eee{\end{equation}}
\def\bef{\begin{figure}}
\def\eef{\end{figure}}
\def\befs{\begin{figure*}}
\def\eefs{\end{figure*}}

\def\OVI {${\rm O\:\scriptstyle VI}$}
\def\HI   {\ifmmode{{\rm H\:\scriptstyle\rm I}}	 \else{${\rm H\:\scriptstyle\rm I}$}\fi}
\def\HII   {${\rm H\:\scriptstyle\rm II}$}
\def\HeI   {${\rm He\:\scriptstyle\rm I}$}
\def\HeII   {${\rm He\:\scriptstyle\rm II}$}
\def\HeIII   {${\rm He\:\scriptstyle\rm III}$}

\def\MgII   {${\rm Mg\:\scriptstyle\rm II}$}

\def\ff   {\ifmmode{f}\else{$f$}\fi}

\def\rmax {\ifmmode{r_{\rm max}}\else{$r_{\rm max}$}\fi}
\def\zmax {\ifmmode{z_{\rm max}}\else{$z_{\rm max}$}\fi}

\def\fCNM {\ifmmode{f^{\rm CNM}}\else{$f^{\rm CNM}$}\fi}
\def\fCMM {\ifmmode{f^{\rm CMM}}\else{$f^{\rm CMM}$}\fi}
\def\fWNM {\ifmmode{f^{\rm WNM}}\else{$f^{\rm WNM}$}\fi}
\def\fWIM {\ifmmode{f^{\rm WIM}}\else{$f^{\rm WIM}$}\fi}
\def\fCM  {\ifmmode{f^{\rm CM}} \else{$f^{\rm CM}$}\fi}
\def\fWM  {\ifmmode{f^{\rm WM}} \else{$f^{\rm WM}$}\fi}

\def\noCNM {\ifmmode{n_o^{\rm CNM}}\else{$n_o^{\rm CNM}$}\fi}
\def\noCMM {\ifmmode{n_o^{\rm CMM}}\else{$n_o^{\rm CMM}$}\fi}
\def\noWNM {\ifmmode{n_o^{\rm WNM}}\else{$n_o^{\rm WNM}$}\fi}
\def\noWIM {\ifmmode{n_o^{\rm WIM}}\else{$n_o^{\rm WIM}$}\fi}
\def\noCM  {\ifmmode{n_o^{\rm CM}} \else{$n_o^{\rm CM}$}\fi}
\def\noWM  {\ifmmode{n_o^{\rm WM}} \else{$n_o^{\rm WM}$}\fi}

\def\nnCNM {\ifmmode{n^{\rm CNM}}\else{$n^{\rm CNM}$}\fi}
\def\nnCMM {\ifmmode{n^{\rm CMM}}\else{$n^{\rm CMM}$}\fi}
\def\nnWNM {\ifmmode{n^{\rm WNM}}\else{$n^{\rm WNM}$}\fi}
\def\nnWIM {\ifmmode{n^{\rm WIM}}\else{$n^{\rm WIM}$}\fi}
\def\nnCM  {\ifmmode{n^{\rm CM}} \else{$n^{\rm CM}$}\fi}
\def\nnWM  {\ifmmode{n^{\rm WM}} \else{$n^{\rm WM}$}\fi}

\def\zCNM {\ifmmode{z_o^{\rm CNM}}\else{$z_o^{\rm CNM}$}\fi}
\def\zCMM {\ifmmode{z_o^{\rm CMM}}\else{$z_o^{\rm CMM}$}\fi}
\def\zWNM {\ifmmode{z_o^{\rm WNM}}\else{$z_o^{\rm WNM}$}\fi}
\def\zWIM {\ifmmode{z_o^{\rm WIM}}\else{$z_o^{\rm WIM}$}\fi}
\def\zCM  {\ifmmode{z_o^{\rm CM}} \else{$z_o^{\rm CM}$}\fi}
\def\zWM  {\ifmmode{z_o^{\rm WM}} \else{$z_o^{\rm WM}$}\fi}

\def\nH {\ifmmode{\langle n_{\rm H} \rangle}\else{$\langle n_{\rm H} \rangle$}\fi}
 
\def \NH {\ifmmode{N_{\rm H}}\else{$N_{\rm H}$}\fi}
\def \NHI {\ifmmode{N_{\rm H}}\else{$N_{\rm H}$}\fi}

\def\Em {\ifmmode{E_m}\else{$E_m$}\fi}
\def\deg   {\ifmmode{^\circ}\else{$^\circ$}\fi}
\def\HH    {\ifmmode{\cal H}\else{${\cal H}$}\fi}
\def\dH    {\ifmmode{\delta\cal H}\else{$\delta{\cal H}$}\fi}
\def\Hdisk {\ifmmode{\cal H}_{\rm disk}\else{${\cal H}_{\rm disk}$}\fi}
\def\Hdiskp {\ifmmode{\cal H}^{'}_{\rm disk}\else{${\cal H}^{'}_{\rm disk}$}\fi}
\def\Hsky  {\ifmmode{\cal H}_{\rm sky}\else{${\cal H}_{\rm sky}$}\fi}

\def\gta{\;\lower 0.5ex\hbox{$\buildrel > \over \sim\ $}}
\def\lta{\;\lower 0.5ex\hbox{$\buildrel < \over \sim\ $}}

\def\Msun   {M$_{\odot}$}

\def\Msun{~$M_{\odot}$}
\newcommand{\Lya}{\mbox {Ly$\alpha$}} 
\newcommand{\Ha}{\mbox {H$\alpha$}}

\def\deg{\hbox{${}^\circ$}}

\def\las{\mathrel{\hbox{\rlap{\hbox{\lower4pt
        \hbox{$\sim$}}}\hbox{$<$}}}}
\def\gas{\mathrel{\hbox{\rlap{\hbox{\lower4pt
            \hbox{$\sim$}}}\hbox{$>$}}}}

\def\arcsec{\hbox{${}^{\prime\prime}$}}

%% file: NH-cutoff-table.tex
\begin{table*}[tbp]
\centering
\caption{\HI\ column density as a function of galactic radius $R$
  (kpc) for different combinations of $J$, $f$, $c$. Column 8: ($J$,
  $f$, $c$) = ($J_o$, 1, 1) is the original model first presented by
  Maloney (1993). The columnar ionization fraction for all models
  increases with radius.  The last line indicates the radius ($r_{0.5}$)
  which corresponds to 50\% ionization of the total hydrogen
  column. The 50\% cut-off radius of Maloney (1993) is emboldened to
  emphasize the radial extent of all other models.}
\label{t:NH}
\begin{tabular}{|l||l|l|l|l|l|l||l|l|l|l|l|l}
\hline
\multicolumn{1}{|c|}{$\log\:$\NH} & \multicolumn{12}{c|}{$R$ (kpc)}
\\ \cline{2-13}  
\multicolumn{1}{|c|}{(cm$^{-2}$)}  & \multicolumn{6}{c|}{$J = J_o$}                                               &
\multicolumn{6}{c|}{$J=0.1 J_o$}
\\ \cline{2-13}  
\multicolumn{1}{|c|}{}
        & $c=1$ & $c=0.1$ & $c=1$ & $c=0.1$ & $c=1$ & $c=0.1$ & $c=1$
& $c=0.1$ & $c=1$ & $c=0.1$ & $c=1$ & \multicolumn{1}{l|}{$c=0.1$}
\\  
        & $f=1$ & $f=1$ & $f=0.1$ & $f=0.1$ & $f=0.01$ & $f=0.01$ &
$f=1$ & $f=1$ & $f=0.1$ & $f=0.1$ & $f=0.01$ &
\multicolumn{1}{l|}{$f=0.01$} \\ \hline 
16                            & 50.9     & 48.7       & 60.2       &
56.9         & 69.5        &  64.9       & 60.2     & 56.9       &
69.5       & 64.9         & 78.8        & \multicolumn{1}{l|}{72.4}
\\ 
16.5                          & 46.5     & 44.8       & 55.5       &
53.0         & 64.8        & 61.1          & 55.5     & 53.0       &
64.8       & 61.1         & 74.0        & \multicolumn{1}{l|}{68.7}
\\ 
17                            & 42.3     & 41.2       & 51.3       &
49.5         & 60.5        & 57.7          & 51.3     & 49.5       &
60.5       & 57.7         & 69.4        & \multicolumn{1}{l|}{65.2}
\\ 
17.5                          & 39.1     & 38.5       & 47.9       &
46.8         & 56.8        & 54.9          & 47.9     & 46.8       &
56.8       & 54.9         & 65.1        & \multicolumn{1}{l|}{61.7}
\\ 
18                            & 37.0     & 36.7       & 45.6       &
44.8         & 53.9        & 52.2          & 45.6     & 44.8       &
53.9       & 52.2         & 61.0        & \multicolumn{1}{l|}{57.4}
\\ 
18.5                          & 35.7     & 35.4       & 43.7       &
42.8         & 50.7        & 48.6          & 43.7     & 42.8       &
50.7       & 48.6         & 55.7        & \multicolumn{1}{l|}{51.8}
\\ 
19                            & 34.2     & 33.8       & 40.9       &
39.7         & 45.8        & 43.3          & 40.9     & 39.7       &
45.8       & 43.3         & 48.5        & \multicolumn{1}{l|}{44.9}
\\ 
19.5                          & 31.7     & 31.1       & 36.1       &
34.7         & 38.5        & 36.4          & 36.1     & 34.7       &
38.5       & 36.4         & 39.4        & \multicolumn{1}{l|}{37.0}
\\ 
20                            & 26.7     & 26.1       & 28.5       &
27.7         & 29.4        & 28.2          & 28.6     & 27.7       &
29.4       & 28.3         & 29.7        & \multicolumn{1}{l|}{28.5}
\\ \hline 
{\bf 50\% ionized}          & {\bf 30.1} & 30.1       & 38.8       & 38.8  & 48.0 &  48.0 &
38.8     & 38.8       & 48.0       & 48.0         & 57.8        & \multicolumn{1}{l|}{57.8}          \\ 
\hline
\end{tabular}
\end{table*}